\documentclass{article}
\usepackage{graphicx}
\usepackage{authblk}
\usepackage{verbatim}
\usepackage{array, tabularx}
\usepackage{subfigure}
\usepackage{caption}

\usepackage[a4paper, total={6.6in, 8in}]{geometry}
\usepackage{xr}
\usepackage[utf8]{inputenc}
\usepackage[T1]{fontenc}
\usepackage{amsthm}
\usepackage{amsmath}
\usepackage{amssymb}
\usepackage{siunitx}

\usepackage{url}
\usepackage{booktabs} % For formal tables
\usepackage{xspace}
\usepackage{float}
\usepackage{multirow}
\usepackage{alltt}
\usepackage{color}
\usepackage{pifont}
\usepackage{makecell}
\usepackage{multicol}
\usepackage{cleveref}
\usepackage[super]{nth}

\title{Fair Voting Outcomes with Impact and Novelty Compromises? Unraveling Biases in Electing Participatory Budgeting Winners}

\author[1]{Sajan Maharjan}
\author[1]{Srijoni Majumdar}
\author[1]{Evangelos Pournaras}

\affil[1]{School of Computing, University of Leeds, Leeds, UK, \ \ \ \ \ \ \ \ \ \ \ \ \ \ \ \ \ \ \ \ \   \ \ \ \ \ \ \ \ \ \ \ \ \ \ \ \ \ \ \ \ \ \ \ \ \   \ \ \ \ E-mails: \{scsmah, s.majumdar,e.pournaras\}@leeds.ac.uk}

\begin{document}
	\maketitle
	
	\begin{abstract}
		\footnotetext[1]{Corresponding author: Sajan Maharjan, School of Computing, University of Leeds, Leeds, UK, E-mail: scsmah@leeds.ac.uk}

Participatory budgeting, as a paradigm for democratic innovations, engages citizens in the distribution of a public budget to projects, which they propose and vote for implementation. So far, voting algorithms have been proposed and studied in social choice literature to elect projects that are popular, while others prioritize on a proportional representation of voters' preferences, for instance, the rule of equal shares. However, the anticipated impact and novelty in the broader society by the winning projects, as selected by different algorithms, remains totally under-explored, lacking both a universal theory of impact for voting and a rigorous unifying framework for impact and novelty assessments. This paper tackles this grand challenge towards new axiomatic foundations for designing effective and fair voting methods. This is via new and striking insights derived from a large-scale analysis of biases over 345 real-world voting outcomes, characterized for the first time by a novel portfolio of impact and novelty metrics. We find strong causal evidence that equal shares comes with impact loss in several infrastructural projects of different cost levels that have been so far over-represented. However, it also comes with a novel, yet over-represented, impact gain in welfare, education and culture. We discuss broader implications of these results and how impact loss can be mitigated at the stage of campaign design and project ideation. 

	\end{abstract}

% \noindent Please note: Abbreviations should be introduced at the first mention in the main text – no abbreviations lists. Suggested structure of main text (not enforced) is provided below.

% \section*{Introduction}

% The Introduction section, of referenced text\cite{Figueredo:2009dg} expands on the background of the work (some overlap with the Abstract is acceptable). The introduction should not include subheadings.

% \section*{Results}

% Up to three levels of \textbf{subheading} are permitted. Subheadings should not be numbered.

% \subsection*{Subsection}

% Example text under a subsection. Bulleted lists may be used where appropriate, e.g.

% \begin{itemize}
% \item First item
% \item Second item
% \end{itemize}

% \subsubsection*{Third-level section}
 
% Topical subheadings are allowed.

% \section*{Discussion}

% The Discussion should be succinct and must not contain subheadings.

%%%%%%%%%%%%%%%%%%%%%%%%%%%%%%%%%%%%%%%%%%%%%%%%%%%%%%%%%%%%%%%%%%%%%%%
\section{Introduction}
\maketitle
Participatory budgeting \cite{wampler_participatory_2021, cabannes2004participatory, aziz2021participatory} is a revolutionary approach to democratic governance, whereby citizens and local governments are actively involved in the decision-making process on the spending of public funds, which are usually carried out via rounds of deliberation alongside voting processes. Since its inception in the city of Porto Alegre, Brazil in 1989 and after more than 11,000 cases worldwide~\cite{pbatlas} with annual investments in the scale of €137M by countries such as Poland~\cite{democracytechnologies}, participatory budgeting initiatives are gaining traction as an effective means to realize local grassroots democratic movements. In a participatory budgeting process~\cite{gomez2013deciding, wampler2000guide, williams2019participatory, bartocci2023journey}, citizens initially propose their project ideas with estimated costs for implementation, given a total budget made available by the city. Initially, participatory budgeting processes involved internal evaluations and selections of proposed projects by the city council after deliberation rounds with citizen representatives. However, by now, recent participatory budgeting designs involve a voting process, where citizens decide on the set of projects to implement. Thus, citizens express their preferences using an input voting method (e.g. approval, cumulative, Knapsack~\cite{goel2019knapsack}) and the final winning projects are calculated using a ballot aggregation method (e.g. utilitarian greedy~\cite{talmon2019framework}, Phragmèn's sequential voting~\cite{brill2023phragmen}, equal shares~\cite{peters2021proportional}). 

\begin{figure}[!htb]
\centering\includegraphics[width=\textwidth]{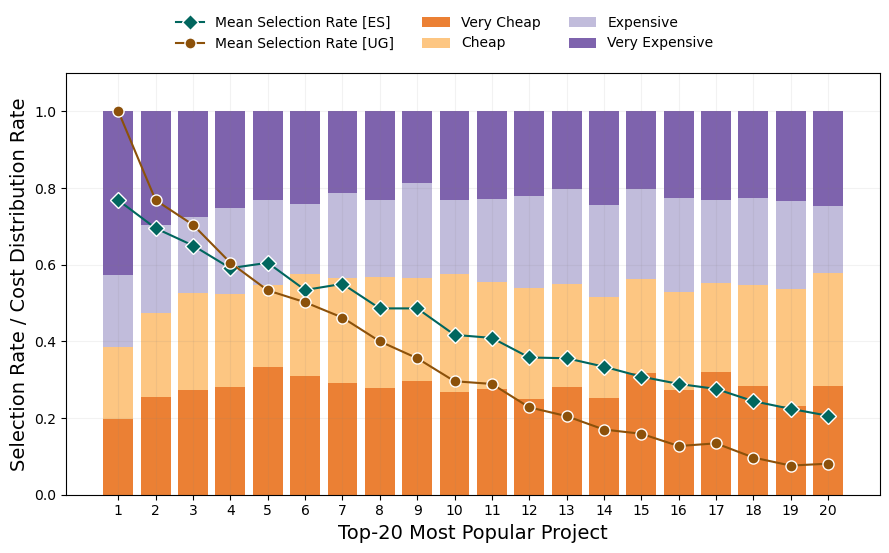}
\caption{\textbf{Equal shares is likely to sacrifice projects among the top-4 popular ones that would otherwise get elected with the utilitarian greedy method, motivating the study of a potential impact loss. Likewise, the most popular projects have a higher share of very expensive projects compared to other cost levels. Nevertheless, compared to utilitarian greedy, equal shares is likely to select (i) more projects that are (ii) less popular.} The plot shows the selection rate (Y axis) of the top-20 most popular projects (X axis) for utilitarian greedy and equal shares over 811 voting instances collected in Pabulib~\cite{faliszewski2023participatory}. These are 613 approval, 103 cumulative and 95 ordinal voting instances counting the scores or votes that the proposed projects receive.}
\label{selection_rate}
\end{figure}

\noindent \textbf{Fair voting outcomes with equal shares}. Recently, the method of equal shares has been tested in real world, for instance, the City Idea project in Aarau, Switzerland~\cite{stadtideeaarau,Welling2023b} or the projects in  
Wieliczka (Green Million) and Swiecie in Poland. This creates a significant momentum for key democratic innovations~\cite{Helbing2015,Helbing2023,Pournaras2020} and a breakthrough for fairer voting outcomes. In contrast to the standard ballot aggregation method of utilitarian greedy that simply selects the next \emph{most popular} project (most received votes) as long as the available budget is not exhausted, the method of equal shares selects projects that aim for a \emph{fairer and more proportional representation} of all voters, by assigning for every voter decision power for an equal part of the budget. The full explanation of axiomatic and algorithmic approaches to fair and proportional voting methods is out of the scope of this paper and it can be found in earlier work~\cite{peters2021proportional, aziz2018proportionally, brill2023proportionality, rey2023computational}. Nevertheless, in practice equal shares results in voting outcomes that may sacrifice a large costly project that can be highly popular (among the ones with the top received votes), to `replace' it with several smaller low-cost projects so that more voters are satisfied (yet, not to the same extent of satisfaction). For instance, in the participatory budgeting campaign of Aarau, the method of equal shares selected 10 more projects than utilitarian greedy (17 vs. 7 out of 33), while strikingly, the third most popular project is sacrificed~\cite{stadtideeaarau,Welling2023b}. Figure~\ref{selection_rate} demonstrates the sacrifice of the top-4 most popular projects by equal shares when applied to earlier empirical evidence of real-world election instances (Pabulib repository data~\cite{faliszewski2023participatory}).

\noindent \textbf{Do fair voting outcomes come with a price}? A large-scale systematic investigation of these particular effects is the motivation and focus of this paper. In particular, we hypothesize that the fairness of equal shares results in potential compromises of the anticipated impact and novelty of the projects in the voting outcomes. This critical inquiry is not made to claim any ineffectiveness or faulty design of equal shares on these aspects, although alternative aggregation methods that optimize for other qualities are plausible. It is also not made to claim that impact loss is anyhow a necessary condition for fairness. Instead, with this inquiry we aim to provide new insights to citizens, city authorities and designers of participatory budgeting campaigns about the proposed projects and their selection to be put for voting such that voting outcomes by equal shares are even more legitimate, impactful and mitigate for any potential (cost) biases, see Section~\ref{discussion}. These insights, missing so far, are critical and timely, and they are expected to significantly strengthen the adoption of methods that promote fairness, such as equal shares. As Cho points out, political fairness is a complex phenomenon, and requires compromise and balancing competing interests so that members of all groups are represented~\cite{cho2018algorithms}. For instance, infrastructure planning and investments to meet net-zero targets in cities can be costly~\cite{deutch2020net, additionalinfranetzero}. To what extent is it likely for equal shares to disadvantage such projects because of their inherently costly nature, resulting in biases affecting sustainability? Or is it likely that multiple such type of projects have more chances to materialize when they have a more localized scope and collectively build, in a more bottom-up way, a sustainability impact at city scale? Do such biases over impact areas also affect in turn the voting turnout within different groups? For instance, Stewart et al. have highlighted the case of participatory budgeting held in Chicago's 49\textsuperscript{th} ward on how a biased voter turnout resulted in the selection of low-priority projects favoring only specific population demographics~\cite{stewart2014participatory}. All these are some key questions that this study puts under scrutiny for the first time and which come with significant policy implications and merit. 

\noindent \textbf{On the challenge of measuring impact of voting outcomes}. Measuring the impact and novelty of voting outcomes is by itself a complex long-standing challenge, and one that this paper addresses. The absence of a standard for impact evaluations on the outcomes of participatory budgeting processes is expected given the inherently ambiguous nature of the term \textit{impact}. Impact is a multi-faceted concept and can be assessed differently by different stakeholders. The European Commission underscores the importance of conducting impact assessments for any public initiatives expected to yield substantial economic, social, or environmental modifications~\cite{europeancommission}. Campbell et al. have conducted a scoping review on the impact of participatory budgeting processes on health and well-being~\cite{campbell2018impact}. Beuermann and Amelina have presented their experimental findings of a participatory budgeting model carried out in Russia, reporting an increased citizens' engagement in public decision-making and raised revenue in the local tax collection~\cite{beuermann2018does}. Hajdarowicz conducts a qualitative analysis on the empowerment of women from participatory budgeting processes~\cite{hajdarowicz2022does}. Cabannes presents case studies of participatory budgeting instances in four cities across Latin America on the initiatives to involve children and young people in the local governance of their cities~\cite{cabannes2006children}. There is a significant body of literature assessing the impact of participatory processes~\cite{mkude2014participatory, wampler2007can, mouter2019introduction,Wellings2023a}, however no existing work evaluates the anticipated impact loss or gain by the winning outcomes of different ballot aggregation methods. This means that our work is distinguished from earlier efforts that focus on measuring the actual impact of the implemented projects. Instead we assess the \emph{anticipated impact} by the projects that constitute the voting outcome of a participatory budgeting process. In this regards, our approach does not replace existing efforts on impact assessment but rather complements them with a novel focus on anticipated impact determined by project selections of different nature. Moreover, distinguishing impact areas of elected participatory budgeting projects is limited to measuring the popularity of impact areas on voters, including how well voters are represented by the voting outcomes~\cite{faliszewski2023participatory,fairstein2022welfare}. Moreover, this earlier work is limited to 4 impact areas and 76 election instances~\cite{faliszewski2023participatory}. Similarly, the recent empirical work by Nelissen on the winning outcomes by equal shares and utilitarian greedy focuses on 35 participatory budgeting instances in Amsterdam~\cite{nelissen2023empirical}. Our study makes a significant advancement in the field by providing stronger validity, an analysis of larger scale and novel universal insights on assessing impact based on new measurements that are not covered in earlier work.

\begin{figure}[!htb]
\centering\includegraphics[width=\textwidth]{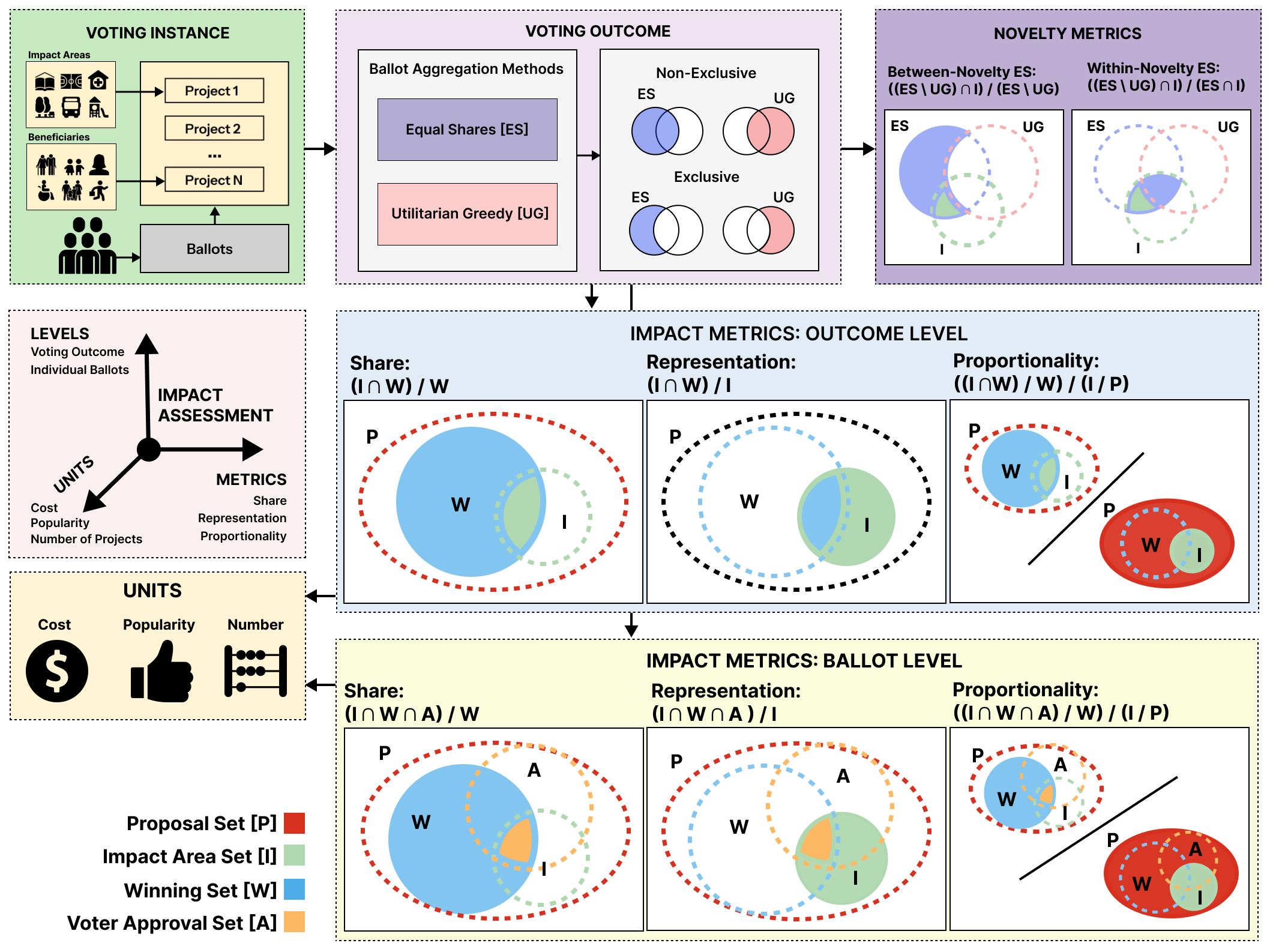}
\caption{\textbf{Impact and novelty assessment framework for voting outcomes by different aggregation methods in participatory budgeting}. For each impact area, the metrics characterize the winning set (share), the proposal set (representation) and both the winning and proposal set (proportionality). These metrics are measured in terms of cost, number of projects or popularity (votes). These metrics are calculated both at the voting outcome and ballot level. The novelty metrics capture the exclusive winning projects by a ballot aggregation method and are distinguished into within-impact-area and between-impact-areas measurements, see Section~\ref{sec:methods} for more information.}
\label{evaluation_framework}
\end{figure}

\noindent \textbf{A framework for measuring impact and novelty}. We introduce a novel framework of impact and novelty measurements on voting outcomes and voters' ballots for participatory budgeting, see Figure~\ref{evaluation_framework}. The framework introduces a number of impact and novelty metrics applied to different impact areas (and project beneficiaries) measured in terms of costs, number of projects and popularity (votes). It consists of the following three elements: (i) The ballot aggregation method that forms the voting outcomes. For this study, we focus on the utilitarian greedy (popularity-oriented) and the equal shares (fairness-oriented) methods. (ii) A number of anticipated impact areas (and beneficiaries) that characterize a proposed project. For instance, a new park with a playground is likely to create impact on the urban greenery and public space development, in particular for children and families. (iii) A number of metrics that characterize the impact areas within the sets of winning projects, proposed projects and both together. \emph{By bringing these three elements together, it becomes possible to assess a number of impact metrics for the voting outcomes of a ballot aggregation method.} We also distinguish for a certain impact area the exclusive winning projects of a particular aggregation method, these are, for instance, the projects of this impact area selected by equal shares but not by utilitarian greedy, and vice versa. Based on this notion of exclusivity, we introduce novelty metrics that can be applied to the number, cost and popularity of projects belonging to an impact area. For both impact and novelty metrics, the loss (or gain) of equal shares over utilitarian greedy is measured as their difference, see Table~\ref{impact_loss_calculation}. 

\noindent \textbf{Impact metrics}. To measure the prevalence of an impact area at the level of a voting outcome or a ballot, we introduce three calculations, each with three units of measurement, making a total of 2 levels x 3 calculations x 3 units of measurement = 18 metrics to assess impact. Figure~\ref{evaluation_framework} shows a visual rationale of these impact calculations using Venn diagrams, while mathematical formulations are given in Section~\ref{sec:methods}. The figures for each impact and novelty measurement can be found for reference in Table~\ref{tracker_impact}. We also introduce a numerical toy example of all impact and novelty calculations in Table~\ref{metric_calculation_example}.

The calculations include the following: (i) \emph{share}--the prevalence of an impact area out of all impact areas in the voting outcome or the ballot. This metric characterizes the winning projects; (ii) \emph{representation}--the prevalence of an impact area in the set of winning projects out of the prevalence of this impact area in the set of proposed projects. This metric characterizes the proposed projects; (iii) \emph{proportionality}--the ratio between the prevalence of an impact area out of all impact areas in the voting outcome or the ballot, and this over the prevalence of this impact area out of all impact areas in the proposed projects. This metric characterizes both the winning and proposed projects.

Each of the three impact calculations measures the prevalence of the impact area in the set of winning or proposed projects as follows: (i) \emph{number of projects}--this counts the projects that belong in the impact area; (ii) \emph{cost}--this is the monetary value of the projects that belong in the impact area; (iii) \emph{popularity}--this is the voters' support on the projects that belong in the impact area. Depending the adopted input voting method, support is counted with the number of approvals or total score assigned to the projects.

\noindent \textbf{Novelty metrics}. These metrics characterize how unique the winning projects of a ballot aggregation method are. Two notions of novelty are distinguished: \emph{within-impact-area novelty} that measures the exclusive winning projects in an impact area by a ballot aggregation method, out of the total \emph{winning projects in this impact area}. In contrast, \emph{between-impact-areas novelty} measures the exclusive winning projects in an impact area by a ballot aggregation method, out of the \emph{exclusive winning projects in all impact areas}. In both cases, unit of measurement can be the number, the cost or the popularity of projects in an impact area. Novelty measurements can be made for both voting outcomes and ballots. The latter reflects the novelty of project choices that voters make in an impact area. Mathematical formulations are given in Section~\ref{sec:methods}. 

\noindent \textbf{Analysis of biases in 345 participatory budgeting elections}. To acquire evidence for any potential impact and novelty loss by the ballot aggregation methods, we apply the proposed framework to the data of 345 out of 810 participatory budgeting elections instances, which are collected from the Pabulib~\cite{faliszewski2023participatory} repository. These participatory budgeting instances are the ones that contain information about the impact areas and beneficiaries of the proposed projects. Each project has one or more labels for the following 9 impact areas: \textit{education, health, welfare, culture, public transit and roads, public space, urban greenery, environmental protection, sport}. Likewise, each project comes with one or more labels for the following 8 beneficiaries: \textit{families with children, students, disabled people, children, adults, animals, youth, elderly}. To control for the factor of cost when assessing the impact and novelty loss, the following additional labels are assigned to each project based on the cost quartile it belongs: \textit{very cheap, cheap, expensive or very expensive}. Controlling for the cost of projects allows us to distinguish between (i) biases on project costs on which the method of equal shares relies for its calculations, and (ii) biases on the project impact area. We also conduct a conjoint analysis~\cite{hainmueller2014causal} to explore causal evidence and explain how the impact areas represented in the winning projects explain the voting outcomes by different ballot aggregation methods. 

\noindent \textbf{Real-world case studies of equal shares for impact loss mitigation}. All the studied participatory budgeting elections used the utilitarian greedy method. It is likely that if another ballot aggregation method was used, the projects and even the choices of voters would be different, for instance, lower-cost project would be proposed, and likely, in different impact areas. To address this threat of validity, we pick up the public voting outcomes~\cite{stadtideeaarau,Welling2023b,equalshareszielony} of two real-world participatory budgeting processes designed to run with the method of equal shares: (i) \emph{Green Million (Wieliczka)} and (ii) \emph{City Idea (Aarau)}. We study these voting outcomes as case studies and hypothesize that they mitigate part of the overall impact and novelty loss. 

\section{Results}\label{results}
\textbf{Four key results} are illustrated in this paper: 
 
 \noindent \textbf{(1)} Equal shares results in voting outcomes with an impact loss in several infrastructural and sustainable development projects, which have shown over-representation. It also results in impact gain in welfare, education and culture, which have been under-represented. Strikingly, impact loss for such infrastructural and sustainable development projects is more frequent, while impact gain in welfare, education and cultural projects is larger in scale. 
 
 \noindent \textbf{(2)} Equal shares results in novelty gain in terms of unique winning projects within each impact area. It also results in novelty gain for the unique winning projects of welfare, education and culture out of all unique winning projects. In all other cases, equal shares shows novelty loss. 
 
 \noindent \textbf{(3)} Equal shares results in impact loss that originates from both high and low cost projects for the impact areas of infrastructural and sustainable development projects. Strikingly, high-cost sport projects show impact loss, while low-cost ones show impact gain. Equal shares results in impact gain for unpopular projects and impact loss for popular and high-cost projects.
 
 \noindent \textbf{(4)} The application of equal shares in two real-world participatory budgeting campaigns mitigates impact loss in public space, urban greenery, sport and, for one of the campaigns, in public transit. Culture and education show lower impact gain than anticipated. 
  
\subsection{Impact loss and gain of equal shares}\label{sacrifice}

\begin{figure}[!htb]
\centering\includegraphics[width=\textwidth]{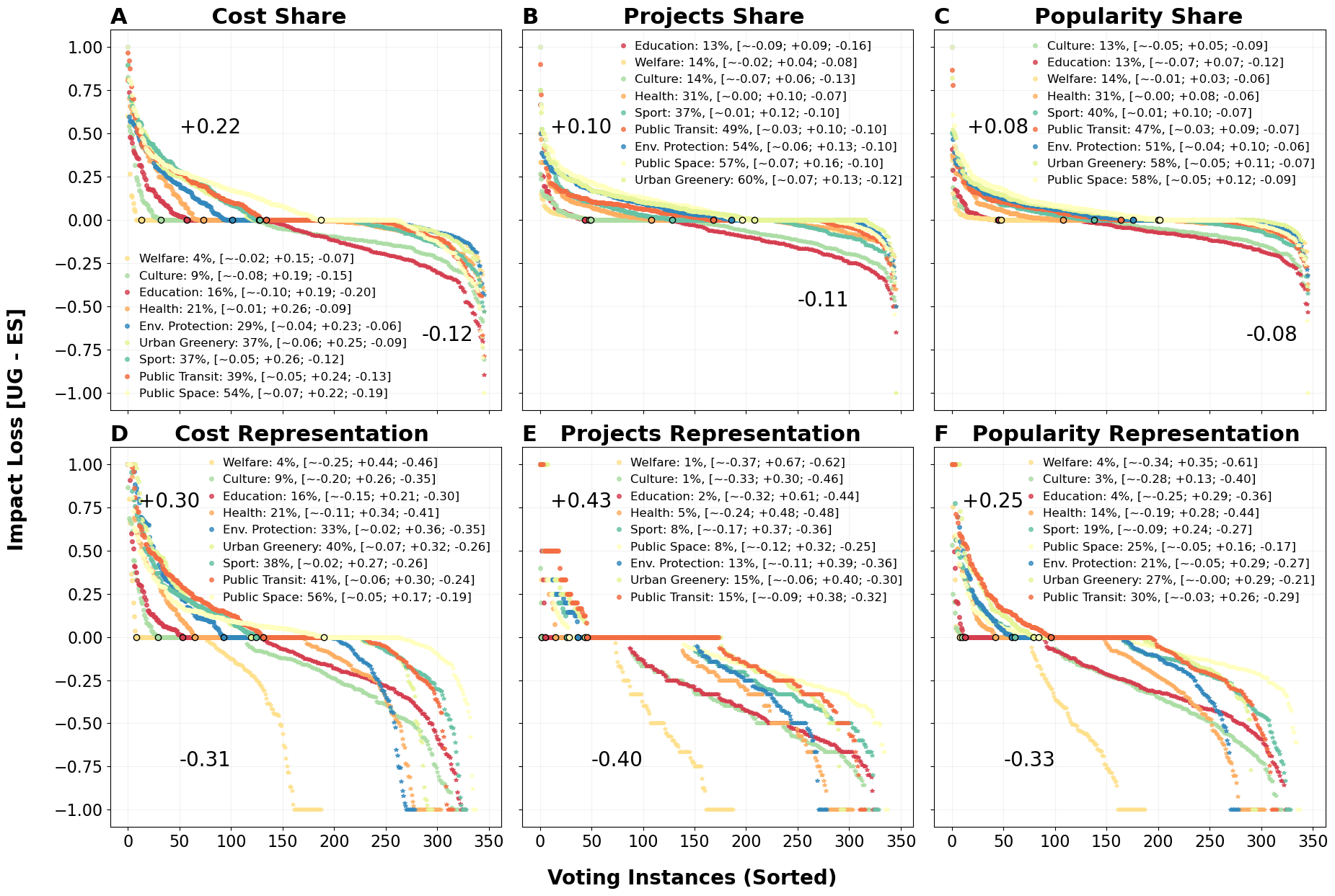}
\caption{\textbf{Equal shares results in voting outcomes with a more frequent impact loss in infrastructural and sustainable development projects, while an impact gain of larger magnitudes is observed in welfare, education and culture projects.} For the depicted metrics, (A) cost share,  (B) projects share, (C) popularity share, (D) cost representation, (E) projects representation and (F) popularity representation, positive loss (Y axis, \textit{UG - ES}) in an impact area (colored lines) shows outperformance by utilitarian greedy (UG), while negative loss shows outperformance by equal shares (ES). The X axis denotes election instances sorted according to impact loss. A circular marker is placed for each line to easier distinguish the number of voting instances with positive/negative loss. The four numbers next to each impact area denote the (i) \% of voting instances with positive loss, (ii) the mean `$\mathtt{\sim}$', (iii) mean positive `+' and (iv) mean negative `-' impact loss. Two additional numbers with the prefixes '+' and '-' placed on each of the metrics signify the overall mean positive and negative impact loss respectively across all impact areas.}
\label{project_type_base_metrics}
\end{figure}

\noindent \textbf{Equal shares has an impact loss in infrastructural and sustainable development projects}. Figure~\ref{project_type_base_metrics} shows the impact loss from utilitarian greedy to equal shares measured with the six metrics distinguished by different impact areas. Projects related to infrastructural and sustainable development such as public space, public transit, urban greenery and environmental protection show higher impact loss. For instance, the highest mean loss of cost share is 7\% for public space, 6\% for urban greenery and 5\% for public transit projects. A similar mean loss is observed for projects and popularity share as well as cost representation, however, the projects and popularity representation show impact gain. This means that although these impact areas are not so prevalent in the winning set, they remain well represented. The voting instances with the cost share losses are 54\%, 36\% and 38\% respectively for public space, urban greenery and public transit. For the metrics of projects and popularity share, these are 57\%, 60\%, 48\% and 58\%, 58\%, 47\% respectively for these impact areas. The voting instances with losses in these impact areas are 56\%, 40\%, 41\% respectively for cost representation, 8\%, 15\%, 14\% for projects representation and 25\%, 27\% and 30\% for popularity representation. 

\noindent \\ \textbf{Equal shares has an impact gain for culture, education and welfare.} Unlike infrastructural and sustainable development projects, there is a net impact gain (as signified by negative values of mean loss) for culture, education and welfare projects across all six impact metrics in Figure~\ref{project_type_base_metrics}. For instance, the mean loss of cost representation is -25\%, -20\% and -15\% for welfare, culture and education. Similar impact gain is observed for projects representation and popularity representation, which is significantly higher than the impact gain for the metrics of cost/projects/popularity share. This means a large portion of the proposed projects in these impact areas are elected. Moreover, the number of voting instances with impact losses is as low as 4\%, 9\%, 16\% respectively for both metrics of cost share and cost representation in such impact areas. Similarly, the voting instances with losses in projects share and projects representation are 14\%, 14\%, 13\% and 1\%, 1\% and 2\% respectively, which is the lowest among other impact areas. The popularity share and popularity representation show a similar pattern.

\begin{figure}[!htb]
\centering\includegraphics[width=0.9\textwidth]{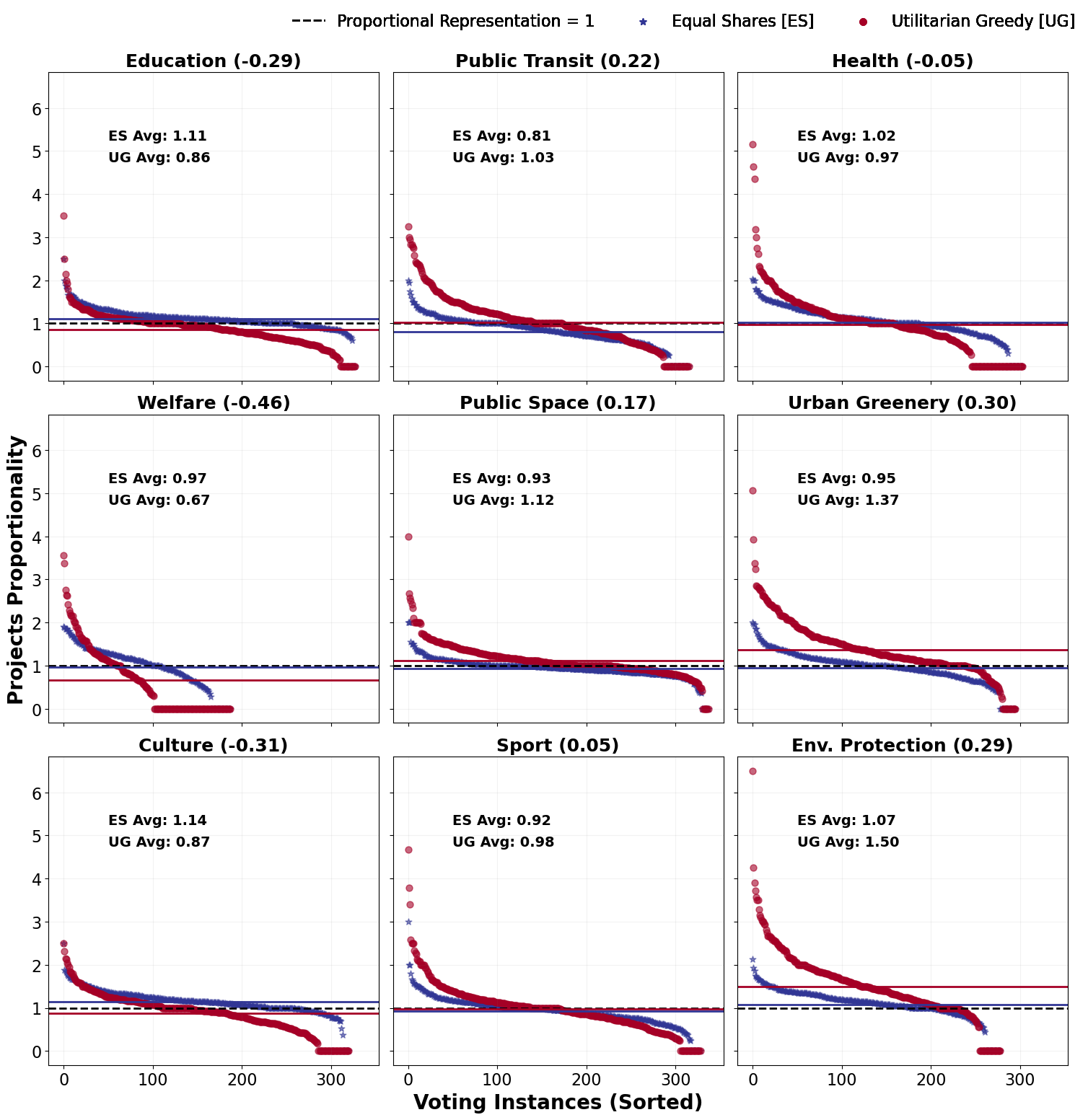}
\caption{\textbf{Equal shares (ES) shows impact gain in projects proportionality for education, welfare, culture and health, with an impact loss in all other impact areas. Equal shares over-represents the impact areas of culture and education, while it under-represents the area of public transit. In contrast, utilitarian greedy (UG) over-represents urban greenery, environmental protection and public space, while under-representing welfare projects.} For each impact area, a projects proportionality value (Y axis) of `1' represents a proportional representation of the impact areas in the winning outcome with respect to the proposed projects. The X axis denotes voting instances sorted according to projects proportionality. The numbers in the parentheses next to each impact area denote the impact loss measured by the mean difference of projects proportionality between utilitarian greedy and equal shares with respect to the value of utilitarian greedy i.e., (UG-ES) / UG.}
\label{project_type_project_representation_ratio}
\end{figure}

\noindent \textbf{Equal shares balances between under-represented and over-represented impact areas.} Figure~\ref{project_type_project_representation_ratio} illustrates the performance of equal shares and utilitarian greedy for projects across different impact areas in terms of projects proportionality. We find that winning outcomes by equal shares are more proportionally represented compared to utilitarian greedy. A value of projects proportionality greater than 1 for an impact area means that the impact area is over-represented in the winning outcome, while a value less than 1 signifies under-representation. Particularly, equal shares increases the proportionality of under-represented impact areas such as education, culture and welfare, while decreases the over-representation of urban greenery and environmental protection. These insights also align with the ones for the cost proportionality (Figure~\ref{cost_proportionality}) and popularity proportionality (Figure~\ref{popularity_proportionality}).

\noindent \textbf{Equal shares better represents project beneficiaries.} The equal shares method also improves the projects representation of any demographic beneficiary. Table~\ref{project_target_win_rates} shows that across all beneficiaries, the projects representation by equal shares increases significantly compared to utilitarian greedy i.e. the winning outcome of equal shares includes projects from which all population groups benefit to higher extent compared to utilitarian greedy. The increase in projects representation for population beneficiaries and a more proportional representation of various impact areas comes at the expense of sacrificing a few larger costly projects for several less costly ones (see Figure~\ref{project_cost_base_metrics}).

\begin{table}[!htb]
\caption{\textbf{Equal shares results in the selection of projects that increases projects representation across all beneficiaries compared to utilitarian greedy.} The table shows the overall number of proposed projects across all 345 instances targeted towards specific beneficiaries. It also shows the corresponding values for projects representation by utilitarian greedy (UG) and equal shares (ES), including the relative loss in representation with respect to utilitarian greedy.}
\label{project_target_win_rates}
\resizebox{1\textwidth}{!}{
\begin{tabular}{lllll}
\hline
Beneficiaries &Total Projects &UG Projects Representation [\%] &ES Projects Representation [\%] &Relative Loss [\%]\\
\hline
Disabled People &4266 &28.3 &61.9 &-119 \\
Youth &6943 &33 &65.5 &-98\\
Seniors &6239 &33.8 &65.7 &-94\\
Children &7062 &33.7 &65 &-93\\
Adults &7656 &35.3 &66.6 &-89\\
Students &366 &44.3 &72.1 &-63\\
Families &2625 &47.6 &66  &-39\\
Animals &313 &67.1 &80.8 &-20\\
\hline
\end{tabular}}
\end{table}

\begin{figure}[!htb]
\centering\includegraphics[width=\textwidth]{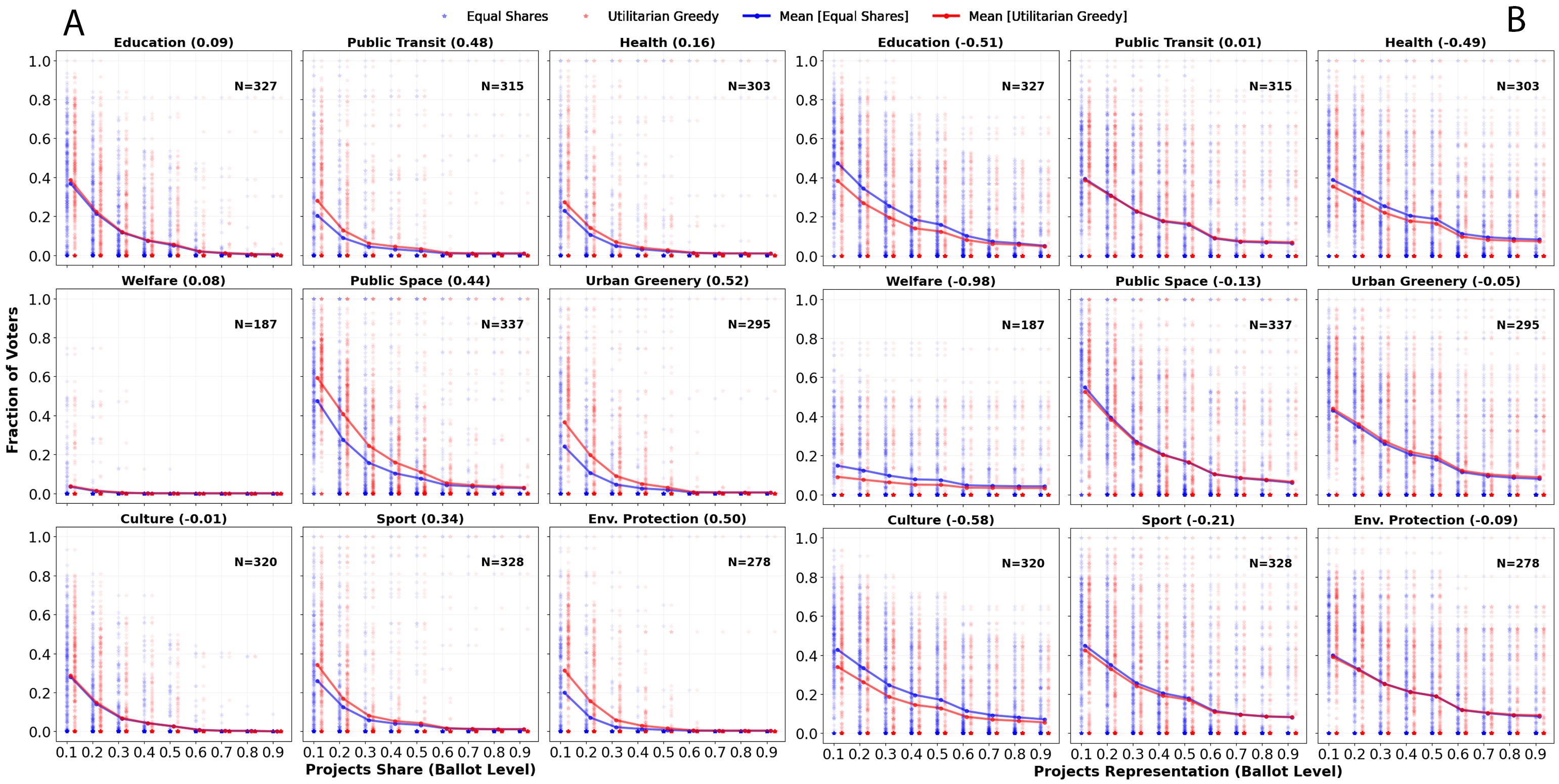}
\caption{\textbf{Equal shares preserves the voters' satisfaction levels for projects with impact gain, while the satisfaction levels are reduced under projects with impact loss. Equal shares shows increased voters' representation levels for projects with impact gain, while the representation levels are not influenced under projects with impact loss.} For each impact area, the fraction of voters (Y axis) for different levels of (A) projects share and (B) projects representation is shown along the X axis for utilitarian greedy and equal shares. A total of 345 approval voting instances are counted, with the value of $N$ denoting the number of election instances with at least one proposed project belonging to the corresponding impact area. The numbers in the parentheses next to each impact area denote the impact loss measured by the mean difference of projects representation at the ballot level between utilitarian greedy and equal shares with respect to the value of utilitarian greedy i.e., \emph{(UG-ES) / UG}.}
\label{fig:impact_area_voter_represenation}
\end{figure}

\noindent \textbf{How impact gains and losses align at the ballot and outcome level} Figure~\ref{fig:impact_area_voter_represenation} illustrates for each impact area (A) the projects share and (B) the projects representation at the ballot level by equal shares and utilitarian greedy. The projects share at the ballot level comes with a loss for equal shares in the impact areas that also show a loss at the outcome level. This is 44\%, 52\%, 50\% and 48\% for public space, urban greenery, environmental protection and public transit. On the other hand, the ones with impact gain at the outcome level, that is education, culture, welfare, and health, come along at the ballot level with higher projects representation gain of 51\%, 58\%, 98\% and 49\% respectively. The difference is particularly prominent for projects representation levels of $<0.6$. In contrast, for the impact areas with an impact loss at the outcome level, such as environmental protection, public space, public transit and urban greenery, representation at the ballot level remains similar for the two ballot aggregation methods.

\subsection{Novelty loss and gain of equal shares} 

\begin{figure}[!htb]
\centering\includegraphics[width=0.9\textwidth]{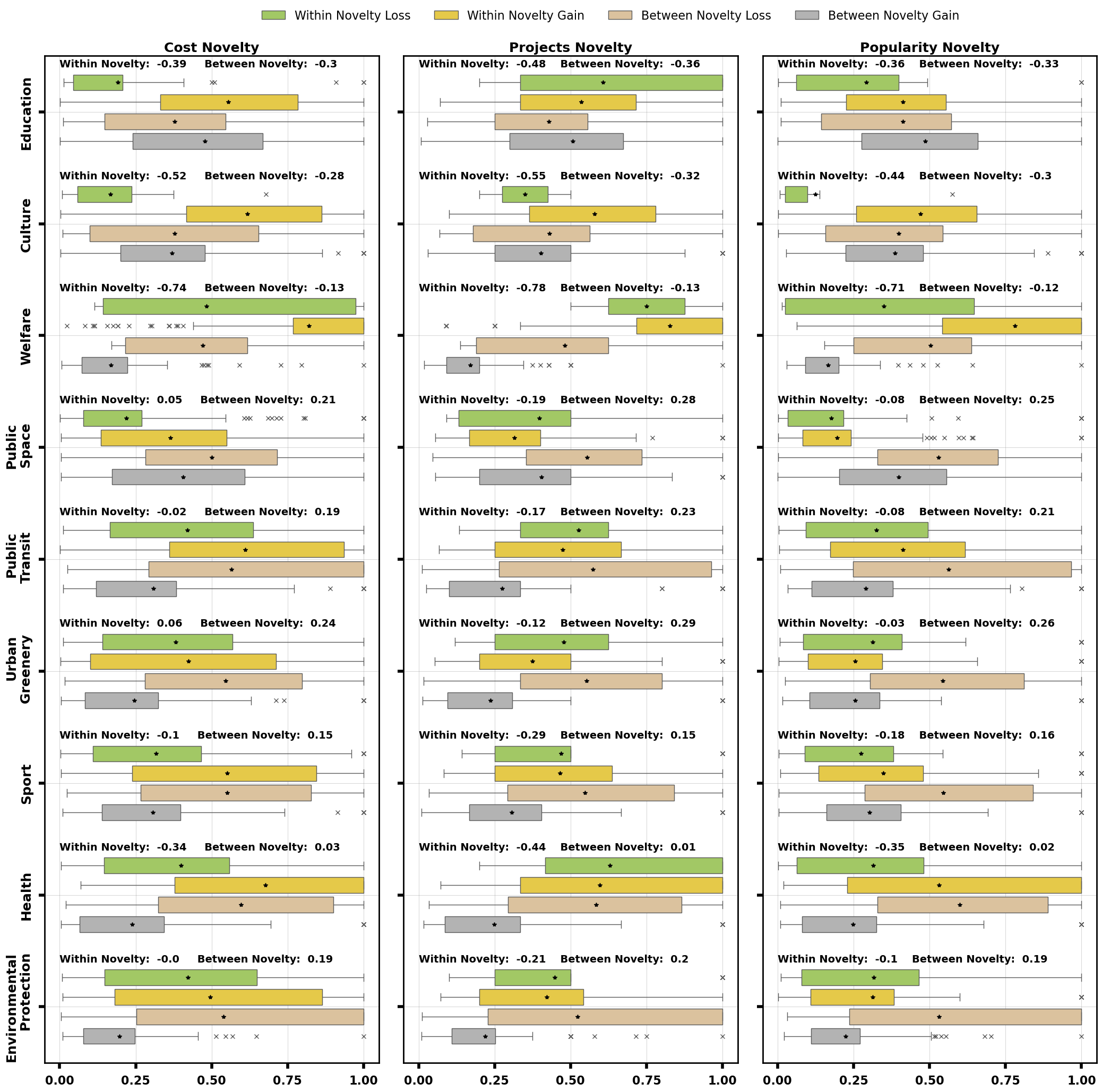}
\caption{\textbf{Equal shares shows within-novelty gain in almost all impact areas and between-novelty gain for education, culture and welfare. In all other impact areas, equal shares shows loss of between novelty. Within novelty comes with larger gains than losses, while between novelty comes with larger losses than gains.} The figure shows different impact areas (Y axis) and the corresponding novelty metric values (X axis). The three columns compare the novelty gain and loss of equal shares over utilitarian greedy in different units: cost, number of projects and popularity.}
\label{project_type_base_metrics_exclusive}
\end{figure}

\noindent \textbf{Novelty gain of equal shares:} Figure~\ref{project_type_base_metrics_exclusive} illustrates the within and between-novelty loss (and gain) across different impact areas. Within-novelty gain is observed across all impact areas. The within-novelty gain is prominent for the impact areas of education, culture and welfare. For instance, in terms of cost and number of projects, the within novelty loss is -39\%, -52\%, -75\% and -48\%, -56\%, -76\% respectively. This means that out of the costs allocated to these impact areas, an additional 39\%, 52\% and 75\% of costs can be attributed to exclusively winning projects by equal shares, which ultimately results in a net impact gain. Equal shares also exhibits between novelty gain but of lower magnitude for the same impact areas of education, culture and welfare i.e. in terms of costs and number of projects, the between novelty losses are -30\%, -28\%, -13\% and  -36\%, -33\% and -13\% respectively. 

\noindent \\ \textbf{Novelty loss of equal shares:} Equal shares shows between-novelty losses for the impact areas of public transit, public space, urban greenery, sport and environmental protection. For instance, in terms of costs, the between-novelty losses for urban greenery, public space, public transit and environmental protection is 24\%, 21\%, 19\% and 19\% respectively. Likewise, in terms of number of projects, the between-novelty losses for the same impact areas are 29\%, 28\%, 23\% and 20\% respectively.

\subsection{Disentangling the interactions of cost, popularity and impact} 

\begin{figure}[!htb]
\centering\includegraphics[width=\textwidth]{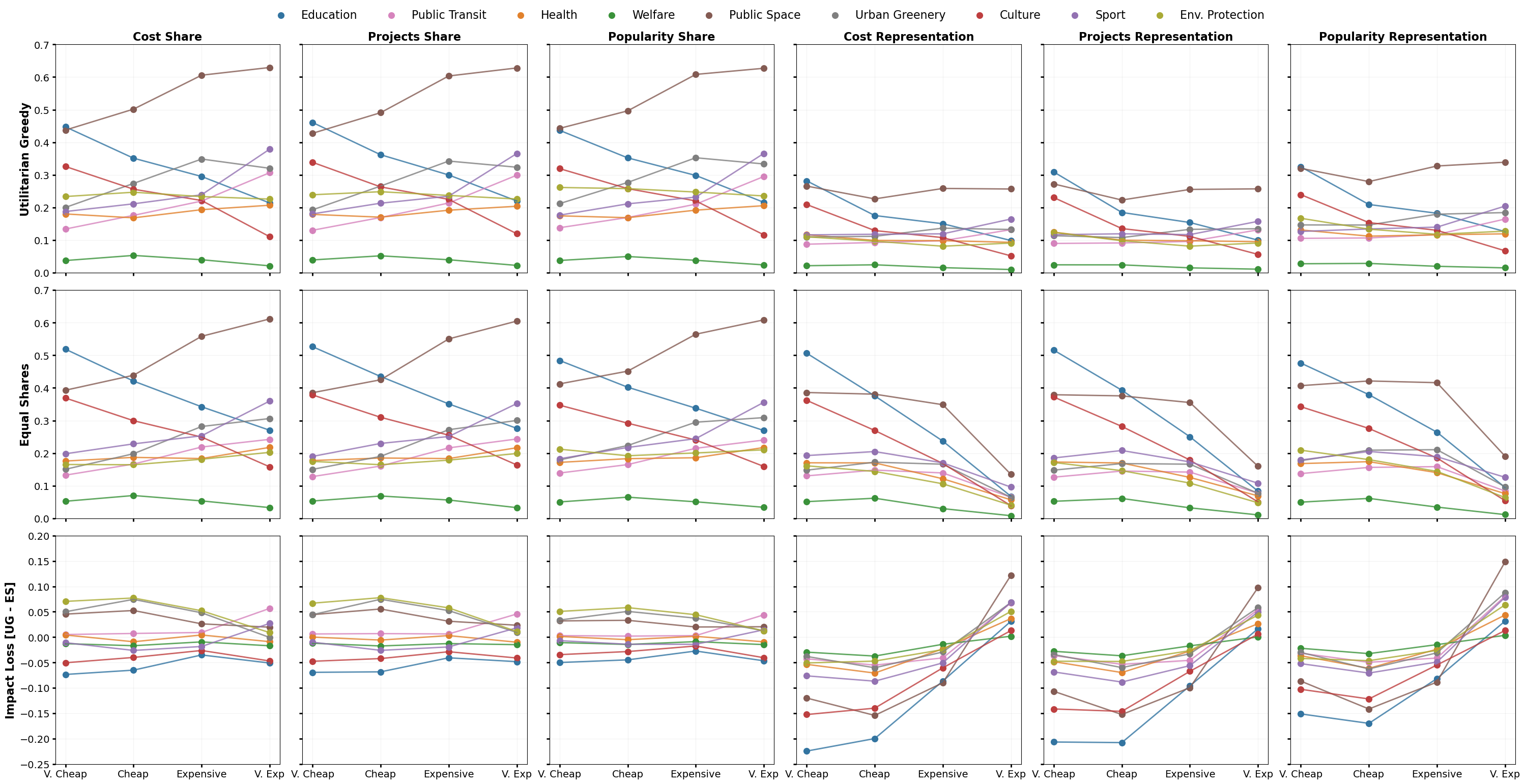}
\caption{\textbf{The impact loss of equal shares for public transit originates from very expensive projects in terms of cost share and projects share. In contrast, the impact loss of equal shares for environmental protection, urban greenery and public space originates from very cheap and cheap projects for the same metrics. Strikingly, very expensive sport projects show impact loss, while cheap and slightly expensive such projects show impact gain. In terms of cost and projects representation, impact loss is observed for very expensive projects but results in impact gain for all other cost levels across all impact areas.} Impact values (Y axis) of equal shares (ES) and utilitarian greedy (UG) across different impact areas for (A) cost share, (B) projects share, (C) cost representation and (D) projects representation segregated at different cost levels (X axis) are shown. The last column depicts the respective impact loss values (\textit{UG - ES}).}
\label{project_types_with_cost}
\end{figure}

\noindent \textbf{Impact gain and loss when controlling for project cost.} To disentangle the interactions between cost of projects and observed impact gain/loss, we control for the cost of the projects. The proposed projects within a single voting instance are classified into four cost levels levels -- \emph{very cheap, cheap, expensive and very expensive} based on the quartile distribution of their costs. Figure~\ref{project_types_with_cost} illustrates the average impact performance of equal shares and utilitarian greedy, and consequently the impact loss or gain of equal shares for projects in different impact areas across these cost levels.

Despite the promotion of low-cost projects by equal shares, very cheap urban greenery projects exhibit an average impact loss of 5\% for cost share and projects share. However, equal shares achieves an impact gain of 4\% for cost and projects representation. For very expensive urban greenery projects, the impact loss of equal shares in cost and projects share is minimal (0\% and 1\% respectively), while loss in cost and projects representation is more significant (7\% and 5\% respectively). Similarly, for very cheap environmental protection projects, the impact loss in cost and projects share by equal shares is 6\% and 7\% respectively. However, equal shares achieves an impact gain of 5\% in cost and projects representation. 

On the other hand, for very cheap educational projects, impact is gained by equal shares for all metrics i.e. cost share, projects share, cost and projects representation by 7\%, 7\%, 23\% and 21\% respectively. For very cheap culture projects, the impact gained for these metrics is 4\%, 4\%, 15\% and 14\% respectively. However, for very expensive educational projects, the impact gain of equal shares is 6\% for cost share and projects share respectively, while a 3\% and 2\% of impact loss is observed in cost and projects representation respectively. Despite the very expensive projects in culture and welfare, equal shares results in an impact gain of 5\%, 2\% in cost share and 4\%, 2\% in projects share respectively. However, impact is lost by a marginal value of 1\% for such very expensive projects in culture and welfare in terms of cost and projects representation.

\noindent \\ \textbf{Impact gain and loss when controlling for popularity.} The impact loss and gain of equal shares is assessed by controlling for different quartile levels of projects popularity: \emph{unpopular, quite popular, popular and very popular}. In the representation metrics, equal shares shows a mean impact gain of 5\% (cost representation), 8\% (projects representation) and 9\% (popularity representation) for unpopular projects but an impact loss of 3\% (projects representation), 3\% (popularity representation) and 8\% (cost representation) for very popular projects (see Figure~\ref{popularity_control_impact_performance}).

\noindent \\ \textbf{How project cost interacts with project popularity.} For the winning projects in utilitarian greedy, the cost share and popularity share show a high correlation of 0.78 , 0.79 and 0.81 (\emph{p-values = $1.01 \times 10 \textsuperscript{-61}$, $4.17 \times 10 \textsuperscript{-76}$, $1.05 \times 10 \textsuperscript{-74}$}) for urban greenery, public space and public transit respectively. However, for such winning outcomes by utilitarian greedy, welfare and culture projects show relatively lower correlations of 0.75 (p-value = $6.98 \times 10 \textsuperscript{-36}$) and 0.69 (p-value = $7.73 \times 10 \textsuperscript{-47}$) that are significant (see Figure \ref{cost_share_vs_popularity_share}).

\noindent \\ \textbf{How impact areas and cost levels explain voting outcomes: a conjoint analysis}. We conduct a conjoint analysis to explore the causal relationship in determining the budget utilization rate of equal shares and utilitarian greedy using the attributes of impact areas and cost levels. Eight independent variables are used (4 variables for popular combinations of impact areas x 2 variables for cost levels). Table~\ref{tags_combination_frequency} shows the most frequent combinations used that cover the range of impact areas. For instance, we observe that two out of four of these combination groups are (i) sustainable infrastructure projects and (ii) culture \& educational projects. For each of these 4 independent variables, we further segregate them into two cost levels -- low-cost and high-cost, resulting in a total of 8 independent variables.

Based on the presence or absence of winning projects with these combinations at two different cost levels in the winning outcomes of equal shares and utilitarian greedy, we model a choice-based conjoint analysis problem to predict the budget utilization (overall cost share). The designed model results in a good fit for both equal shares and utilitarian greedy with $R^2=0.88$ and $R^2=0.82$ respectively and significant p-values for all independent variables i.e. \emph{p-value < 0.05} (see Table~\ref{conjoint_analysis_results}).

\begin{figure}[!htb]
\centering\includegraphics[width=0.9\textwidth]{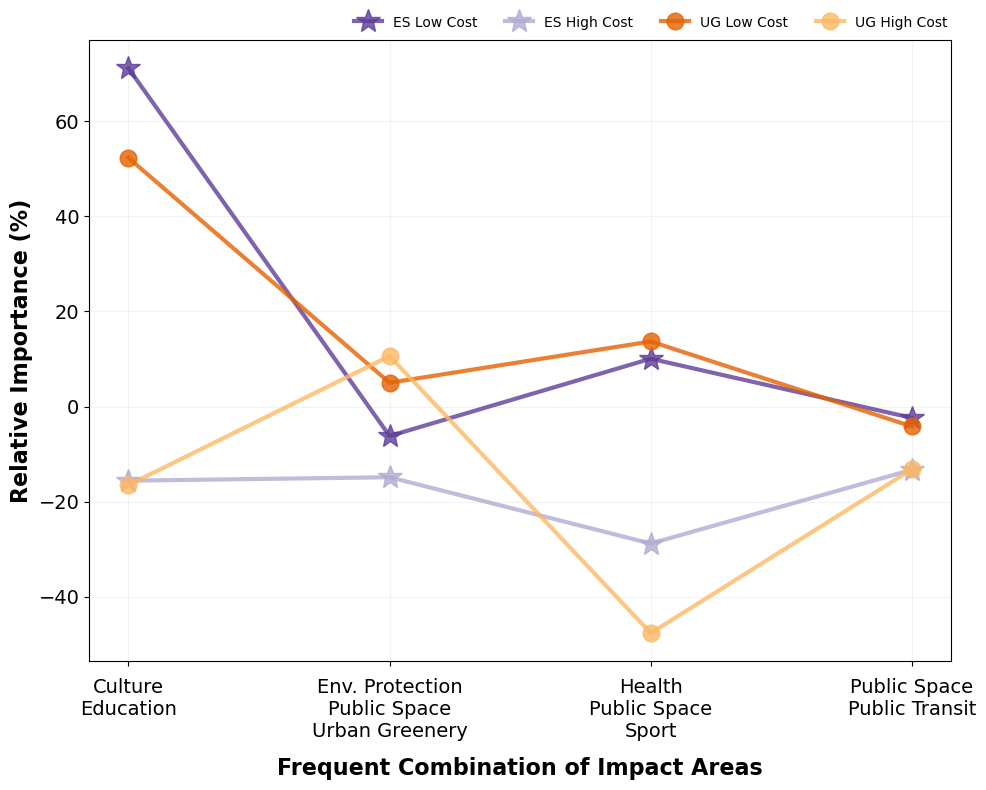}
\caption{\textbf{The impact areas of culture and education with low cost have the highest importance in both equal shares and utilitarian greedy, while for high costs, both of these impact areas come with negative importance. This pattern is also observed for the combinations of health, public space, sport and public space, public transit. In contrast, for the impact area of environmental protection, public space and urban greenery, the ballot aggregation method is more important than the cost: positive importance for both cost levels of utilitarian greedy, negative importance for both cost levels of equal shares}. The most frequent combinations of impact areas (X axis) are shown across cost levels and their relative importance (Y axis, part-worth utilities) contributing to the \emph{budget utilization} by equal shares and utilitarian greedy.} 
\label{relative_importance_conjoint_analysis}
\end{figure}

The relative importance (part-worth utilities) of different impact areas and cost levels by equal shares and utilitarian greedy is shown in Figure~\ref{relative_importance_conjoint_analysis}. The conjoint analysis further reinforces the findings of our results. The figure illustrates that projects with the combination of \textbf{\textit{education, culture}} at low-cost contribute the most to budget utilization by both equal shares and utilitarian greedy (relative importance of 71.6\% and 52.5\% respectively). However, the same combination of projects at higher cost levels contributes negatively to the budget utilization rate in both aggregation methods. Likewise, projects with combinations of environmental protection, public space and urban greenery across both cost levels contribute positively to the budget utilization in the case of utilitarian greedy, whereas such projects contribute negatively in the case of equal shares. This observation is congruent with results that suggest impact is lost by equal shares for sustainable infrastructure projects, while impact is gained for projects related to culture, education and welfare.

\subsection{Equal shares in real world: the cases of City Idea \& Green Million}
\begin{figure}[!htb]
\centering\includegraphics[width=\textwidth]{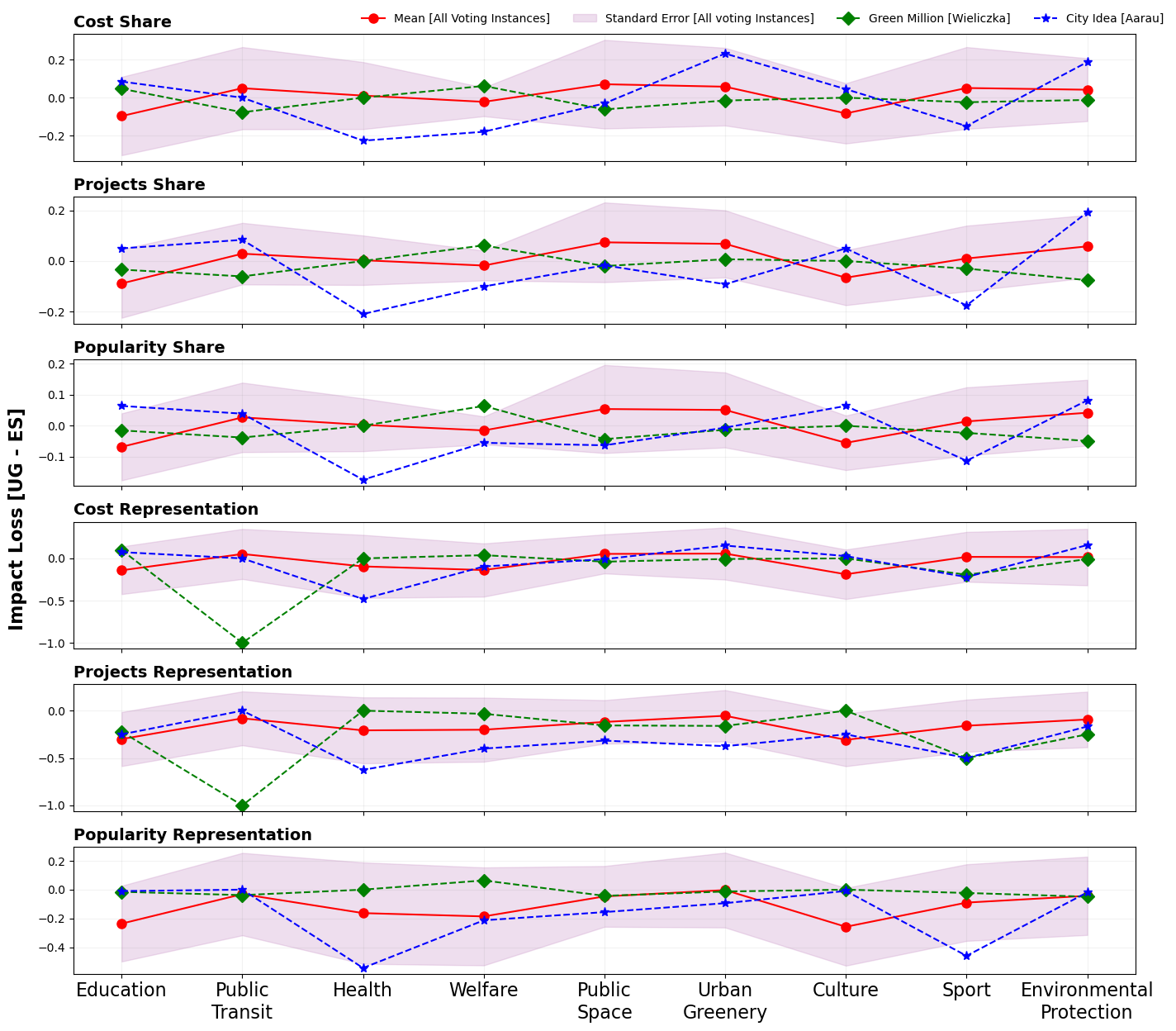}
\caption{\textbf{The application of equal shares in real world mitigates expected impact loss in public space and urban greenery for cost share and projects share, while impact loss is observed in culture and education and impact gain in sport. A significant impact gain for public transit is observed in the Green Million campaign in Poland. Welfare in the City Idea campaign in Switzerland shows impact gain for cost share and projects share.} The impact loss (Y axis) for different impact metrics across different impact areas (X axis) for (i) all 345 voting instances (mean), (ii) the Green Million campaign in Wieliczka, Poland and (iii) the City Idea campaign in Aarau, Switzerland is shown.}
\label{city_idea_zielony}
\end{figure}

\noindent \textbf{Does equal shares mitigate impact loss in real world?} Here we explore the impact loss and gain of equal shares across different impact areas for the recently implemented participatory budgeting instances in Aarau (City Idea Project) and Wieliczka (Green Million), in which winning outcomes were determined using a variant of the equal shares method (\emph{add1} using integral endowments for Green Million and \emph{add1u} for City Idea~\cite{pabutools}). The impact loss (or gain) on the winning outcomes by equal shares for these two cases along with the mean values of all voting instances is shown in Figure~\ref{city_idea_zielony}.

The winning outcomes of equal shares for the Green Million participatory budgeting instance are different from the expected winning outcomes of equal shares based on the past voting instances. Projects under the impact areas of public transit, public space, urban greenery and environmental protection achieve an additional impact gain instead of the expected impact loss for such projects. For instance, public space projects achieve an impact gain of 6\% in cost share instead of an expected impact loss of 7\%. Likewise, environmental protection projects show an impact gain of 8\% in projects share instead of an expected impact loss of 6\%. This is because the Green Milion Project, as its name suggests, was particularly aimed at promoting eco-friendly "green" projects such that most of the proposed projects are related to environmental protection, urban greenery and public space. Likewise, due to the absence of projects related to education, culture and welfare for this campaign, an unexpected and significant impact loss is observed for such impact areas in all metrics. For instance, educational projects have a cost share loss of 5\% rather than the expected gain of 10\%, while welfare projects have a projects share loss of 6\%. In case of the City Idea campaign in Aarau, the winning outcomes by equal shares exhibit the expected impact gains for welfare (an average of 17\% across all metrics) and the expected impact loss for public transit (8\% projects share loss, 4\% popularity share loss) and environmental protection (18\% cost share loss, 19\% projects share loss, 8\% popularity share loss, 16\% cost representation loss). However, unlike the expected impact gains of equal shares for projects related to education, a significant impact loss for educational projects (8\% cost share loss, 5\% projects share loss, 7\% projects representation loss) is observed. Moreover, a significant impact gain is observed for projects related to urban greenery (9\% projects share gain, 9\% popularity representation gain).

\section{Discussion}\label{discussion}

The findings and key results comes with some significant implications. Equal shares gains momentum with radically different voting outcomes in participatory budgeting: a larger number of lower cost projects are elected (see Figure~\ref{project_cost_base_metrics}), often `replacing' expensive popular projects, creating a fundamentally different impact for the society as a result of higher proportional representation of voters' preferences. We show that there are some opportunities and risks here. In particular, infrastructural projects, especially ones related to sustainability, may be frequently disadvantaged by equal shares, as a result of their inherently costly nature but also beyond this, see Figure~\ref{project_types_with_cost}. On the other hand, equal shares strongly favors novel welfare, education and cultural projects, which is a likely result of their lower cost in the context of participatory budgeting. 

The risk of a democratic deficit towards the endeavor of sustainability and net zero requires attention and mitigation. Apparently, letting infrastructural and sustainability projects planned in a top-down way is a major threat to legitimacy~\cite{Hausladen2023} by creating vulnerabilities for corruption and even green washing by large corporate organizations with influence in governments. Preserving a capacity in participatory budgeting campaigns to materialize infrastructural and sustainability projects that fairly benefit the whole population is a key challenge to address in the future. The new findings here suggest that participatory budgeting campaigns with projects of high discrepancies of cost may be more effective when they are broken down into multiple campaigns with more local and synergistic effects in their implementation across different localities, for instance, low-cost incremental expansions of bike lanes across a city, instead of centrally deciding the creation of a large multi-million bike lane network. The possibility of such substitution or complementarity effects among proposed projects is also discussed by Jain et al.~\cite{jain2020participatory}. Furthermore, the insights of this study are also pertinent to national elections when considering the political agendas of different candidates and in which impact areas their priorities lie, for instance, climate change and social welfare.

We also show evidence from the first recent real-world participatory budgeting campaigns conducted using the method of equal shares that the mitigation of this impact loss is feasible. Awareness about how the ballot aggregation method works and what it prioritizes results in different proposed projects, with different (lower) costs and, eventually, different preferences that yield impact recovery for public transit and sport, see Figure~\ref{city_idea_zielony}. 

This study provides new insights for several beneficiaries: For \emph{citizens and communities}, the results unravel new pathways for democratic social innovations and the preservation of capacity to innovate as such. They also provide further empirical evidence and understanding of more complex voting methods such as equal shares, and its effectiveness. For \emph{policymakers}, the findings of this work can support a more effective design of participatory budgeting campaigns that mitigate for potential impact losses already at the stage of project proposals and feasibility checks. They also support them to align ambitious net zero policy agendas with citizens' participation in their implementation and democratic legitimacy~\cite{Wellings2023a}. For the \emph{academic community} of (computational) social choice and beyond, the insights of our study motivate for new axiomatic foundations to assess fairness, proportionality and voting methods. Last but not least, for \emph{industry}, a more localized, distributed and synergistic implementation of infrastructural and sustainability projects with direct citizens' engagements suggests the need for new business models aligned to democratic values. 

\section{Methods}\label{sec:methods}

This section outlines the empirical dataset used to apply the impact assessment framework as well as the mathematical formulations of all impact and novelty metrics. We also outline how we account for threats to validity. Table~\ref{tracker_impact} outlines the metrics and the figures that illustrate the respective impact assessment results.

\begin{table}[!htb]
\centering
% \captionsetup{justification=raggedright, singlelinecheck=false}
\caption{Metrics [Impact \& Novelty] - Results Mapping}
\label{tracker_impact}
% \resizebox{0.88\textwidth}{!}{%
    \begin{tabular}{lllllll}
    \toprule
    % \multicolumn{}{}{}
    \multirow{2}{*}{Dimension} & \multicolumn{2}{c}{\# of Projects}  & \multicolumn{2}{c}{Cost}& \multicolumn{1}{c}{ Popularity} \\
    \cmidrule(lr){2-6}
          & Outcome & Individual Ballot & Outcome & Individual Ballot & Outcome \\
    \midrule
    
    Share &	Figure 3 &	Figure 5 &	Figure 3	&	Figure S7 & Figure 3\\ 
    Representation &	Figure 3 &	Figure 5 &	Figure 3	& Figure S8	& Figure 3\\ 
    Proportionality&	Figure 4 &	Figure S9 &	Figure S5	& Figure S10	& Figure S6\\ 
    \midrule
    \multicolumn{6}{c}{Between-Novelty and Within-Novelty Impact Areas = Figure 6}\\
    \bottomrule
    \end{tabular}%
% }
\end{table}

\subsection{Empirical Dataset}

The data on participatory budgeting elections was collected in August 2023 from the Pabulib~\cite{faliszewski2023participatory} open repository. The repository consists of multiple files which signify the details of a particular participatory budgeting voting instance defined by a standard .pb file. Such files contain details on project costs, votes received, project description, and the total available budget. In reality, for all except two of the voting instances from the collected dataset, the winning outcomes were determined using utilitarian greedy. To determine the hypothesized winning outcomes by equal shares, we used the pabutools~\cite{pabutools} library. While calculating the winning outcomes by equal shares, we added the conditions of arbitrary budget increments and utilitarian termination (\emph{add1u}~\cite{pabutools}) to ensure high utilization of budget.

The dataset consists of 810 participatory budgeting instances. Out of these, only 345 instances have information on project impact areas and population beneficiaries. Table~\ref{project_impact_area_beneficiaries_distribution} shows the overall and average distribution of projects across the 9 different impact areas and 8 different beneficiaries. The distribution of number of projects, their costs as well as their popularity across different impact areas is shown in Figure~\ref{project_distribution}. Table~\ref{tags_combination_frequency} shows the most frequent combination of project impact areas from the observed dataset used in the conjoint analysis. 

% Tabular data for count distribution of project types
\begin{table}[!htb]
\centering
\caption{\textbf{Distribution of proposed projects across different impact areas and beneficiaries}}
\label{project_impact_area_beneficiaries_distribution}
\begin{tabular}{llll}
\toprule
Impact Areas / Beneficiaries & Total Projects & Avg. Num. Projects \\
\midrule
Education &4336 &13.3  \\
Health &1408 &4.6 \\
Culture &3019 &9.4 \\
Sport &2760 &8.4 \\
Public Transit &1973 &6.3\\
Welfare &783 &4.2 \\
Urban Greenery &2534 &8.6  \\
Public Space &5512 &16.4 \\
Environmental Protection &2093 &7.5 \\
\midrule
Families &2625 &7.7 \\
Students &366 &1.1 \\
Disabled People &4266 &12.4 \\
Children &7062 &20.5 \\
Adults &7656 &22.3\\
Seniors &6239 &18.1\\
Animals &313 &0.9\\
Youth &6943 &20.2\\
\bottomrule
\end{tabular}
\end{table}

\begin{table}[htb!]
\centering
\caption{\textbf{Most frequently observed mutually-exclusive combination of labels for project impact areas}}
\label{tags_combination_frequency}
\begin{tabular}{ll}
\toprule
Project Tags Combination &Number of Projects \\
\midrule
Culture, Education &1227\\
Environmental Protection, Public Space, Urban Greenery &966 \\
Public Space, Public Transit and Roads &755 \\
... &... \\
Health, Public Space, Sport &261\\
\bottomrule
\end{tabular}
\end{table}

\subsection{Impact and Novelty Metrics}
Table~\ref{list_of_symbols} outlines the list of mathematical symbols and their meanings, which define the impact and novelty metrics.  They are assessed on the voting outcomes and the individual voters' ballots. In the following sections, we rigourously introduce how the impact and novelty metrics are calculated at the voting outcome level and the ballot level. For both the impact and novelty metrics, the loss (or gain) value of equal shares relative to the utilitarian greedy approach is quantified as the difference between the two values, as presented  in Table~\ref{impact_loss_calculation}. Furthermore, a toy voting example with 11 voters and 3 different impact areas, shows how the proposed impact and novelty metrics are calculated (see Tables~\ref{toy_voting_example} and~\ref{metric_calculation_example}).

\begin{table}[!htb]
\caption{\textbf { {List of mathematical symbols }}}
\label{list_of_symbols}
\resizebox{1\textwidth}{!}{
    \begin{tabular}{ll}
    \toprule
    Symbol &Interpretation \\
    \midrule
    $o$ &voting outcome\\
    $b$ &individual ballot\\
    $l$ & an impact area\\
    $L$ & set of impact areas\\
    $p$ & a proposed project\\
    $P$ & set of proposed projects \\
    ${c}_{p}$ & cost of a project ${p}$, where ${p}$ $\in$ $P$\\
    $P_l$ & set of proposed projects belonging to an impact area $l$, where $l\ \in\ L$ \\
    $V$ & set of voters\\
    $B_v$ & set of approved projects for a voter $v \in V$\\
    ${v}_{p}$ & votes received by a project ${p}$, where ${p}$ $\in$ $P$\\
    $\mathsf{u}$ &  utilitarian greedy \\
    $\mathsf{e}$ &  equal shares \\
    $f$ & ballot aggregation method i.e. $f = \{e, u\}$ \\
    $W_f$ & set of proposed projects that are winners in the outcome of an aggregation method $f$ \\
    $\widehat{W}_{{f}}$ & set of proposed projects that are exclusive winners in the outcome of an aggregation method $f$ \\
    ${W_{l,f}}$  & $P_l\  \cap W_f$  \\
    $\widehat{W}_{{l,f}}$  & $P_l\  \cap \widehat{W}_{{f}}$  \\
    ${W_{l,v,f}}$  & $P_l\ \cap B_v\ \cap W_f$  \\
    $\widehat{W}_{{l,v,f}}$  & $P_l\ \cap B_v\ \cap\ \widehat{W}_{{f}}$  \\
    $r_{c, l}$ & ratio of total cost of projects in the impact area $l$ over the total cost of proposed projects, i.e. $\frac{\sum_{p=1}^{|{P_{l}}|} {c}_{p}}{\sum_{p=1}^{|P|} {c}_{p}}$\\
    $r_{n,l}$ & ratio of number of proposed projects in the impact area $l$ over the total number of proposed projects, i.e.  $\frac{|{P_{l}|} }{{|P|}}$ \\
    $r_{v, l}$ & ratio of total votes received by projects in the impact area $l$ over the total votes received by proposed projects, i.e $\frac{\sum_{p=1}^{|{P_{l}}|} {v}_{p}}{\sum_{p=1}^{|P|} {v}_{p}}$\\
    $i$ & impact metric \\
    $\omega$ & within-novelty metric \\
    $\beta$ & between-novelty metric \\
    \bottomrule
\end{tabular}
}
\end{table}

\noindent \\ \textbf{Impact metrics calculated at the voting outcome level:}
The prevalence of the impact areas is measured in terms of their share, representation and proportionality in the voting outcome $o$  of an aggregation method $f$. For different voting aggregation methods, each of the impact metrics is measured in terms of (a) cost, (b) number of projects and (c) popularity to assess the impact on the voting outcome.

{\em Share}: This metric quantifies the winning rate of projects from a specific impact area with respect to the winning outcome. For an aggregation method $f$, share is calculated as the fraction of the total cost, number, or popularity of projects in the winning outcome that belongs to a given impact area $l$:

\begin{table}[H]
\captionsetup{labelformat=empty}
\label{w}
\centering
\begin{tabular}{lll}

Cost Share & Projects Share & Popularity Share \\
\midrule
 ${S}_{o,c,l}^{i,f}$  = $\frac{\sum_{p=1}^{|{W_{l,f}}|} {c}_{p}}{\sum_{p=1}^{|{W_{f}}|} {c}_{p}}$ & ${S}_{o,n,l}^{i,f}$  = $\frac{|W_{l,f}|}{|W_{f}|}$ & ${S}_{o,v,l}^{i,f}$  = $\frac{\sum_{p=1}^{|{W_{l,f}}|} {v}_{p}}{\sum_{p=1}^{|{W_{f}}|} {v}_{p}}$ \\
\end{tabular}
\end{table}
{\em Representation}: This metric quantifies the rate of a specific impact area represented in the winning set with respect to the set of proposed projects in that impact area. For an aggregation method $f$, representation is calculated as the fraction of the total cost, number, or popularity of projects in a given impact area, $l$, that belongs to the winning set:

\begin{table}[H]
\captionsetup{labelformat=empty}
\label{w}
\centering
\begin{tabular}{lll}
Cost Representation  & Projects Representation  & Popularity Representation \\ 
\midrule
${R}_{o,c,l}^{i,f}$  = $\frac{\sum_{p=1}^{|{W_{l,f}}|} {c}_{p}}{\sum_{p=1}^{|{P_{l}}|} {c}_{p}}$ & ${R}_{o,n,l}^{i,f}$  = $\frac{|{W_{l,f}}|}{|{P_{l}}|}$ &  ${R}_{o,v,l}^{i,f}$  = $\frac{\sum_{p=1}^{|{W_{l,f}}|} {v}_{p}}{\sum_{p=1}^{|{P_{l}}|} {v}_{p}}$ \\
\end{tabular}
\end{table}

{\em Proportionality}: This metric characterizes the proportional representation of impact areas across both the winning and proposal sets. Given that a certain impact area constitutes a specific fraction out of all proposed projects, the metric signifies if a proportional fraction is maintained in the winning outcome for that impact area. For an aggregation method $f$, proportionality is calculated as the ratio of winning shares for an impact area $l$ over the initial fraction of that impact area across all proposed projects:  

\begin{table}[H]
\captionsetup{labelformat=empty}
\label{w}
\centering
\begin{tabular}{lll}

Cost Proportionality  & Projects Proportionality  & Popularity Proportionality \\ 
\midrule
$\textit{P}_{o,c,l}^{i,f}$ = $\frac{{S}_{o,c,l}^{i,f}}{{r}_{c,l}}$ & $\textit{P}_{o,n,l}^{i,f}$ = $\frac{{S}_{o,n,l}^{i,f}}{{r}_{n,l}}$ & $\textit{P}_{o,v,l}^{i,f}$ = $\frac{{S}_{o,v,l}^{i,f}}{{r}_{v,l}}$ \\
\end{tabular}
\end{table}

\noindent \textbf{Impact metrics calculated at the ballot level} In this case, share, representation and proportionality metrics across different impact areas are computed for each voter to understand how voters' choice relate to the voting outcomes, and how impact gain or loss aligns at the individual voters' ballots and outcome level. \emph{Popularity is not accounted at the individual ballot level, because popularity is formulated from the collective set of individual ballots.}

{\em Share}:  For an aggregation method $f$, share at the ballot level is calculated as the fraction of the total cost, number, or popularity of projects in the winning outcome that belongs to a given impact area $l$ which is also approved by voter $v$:

\begin{table}[!htb]
\captionsetup{labelformat=empty}
\label{w}
\centering
\begin{tabular}{ll}

Cost Share  & Projects Share  \\
\midrule
${S}_{b,c,l}^{i,f}$  = $\frac{\sum_{p=1}^{|{{W}_{l,v,f}}|} {c}_{p}}{\sum_{p=1}^{|{{W}_{f}}|} {c}_{p}}$ & ${S}_{b,n,l}^{i,f}$  = $\frac{|{{W}_{l,v,f}}|}{|{{W}_{f}}|}$ \\

\end{tabular}
\end{table}

{\em Representation}:  For an aggregation method $f$, representation at the ballot level is calculated as the fraction of the total cost, number, or popularity of projects in a given impact area, $l$, that belongs to the winning set which is also approved by the voter $v$:

\begin{table}[!htb]
\captionsetup{labelformat=empty}
\label{w}
\centering
\begin{tabular}{ll}

Cost Representation  & Projects Representation  \\
\midrule
${R}_{b,c,l}^{i,f}$  = $\frac{\sum_{p=1}^{|{{W}_{l,v,f}}|} {c}_{p}}{\sum_{p=1}^{|{{P}_{l}}|} {c}_{p}}$ & ${R}_{b,n,l}^{i,f}$  = $\frac{|{W}_{l,v,f}|}{|{P}_{l}|}$ \\
\end{tabular}
\end{table}

{\em Proportionality}: For an aggregation method $f$, proportionality at the ballot level is calculated as the ratio of winning shares for an impact area $l$ which is also approved by voter $v$ over the initial fraction of that impact area across all proposed projects:

\begin{table}[H]
\captionsetup{labelformat=empty}
\label{w}
\centering
\begin{tabular}{ll}

Cost Proportionality  & Projects Proportionality  \\
\midrule
${P}_{b,c,l}^{i,f}$  = $\frac{{S}_{b,c,l}^{i,f}}{{r}_{c,l}}$ & ${P}_{b,n,l}^{i,f}$  = $\frac{{S}_{b,n,l}^{i,f}}{{r}_{n,l}}$  \\
\end{tabular}
\end{table}

\noindent \\ \textbf{ {\em Within-novelty} metrics calculated at the voting outcome level}: The novelty $\omega$ within an impact area $l$ for an aggregation method $f$ is calculated as the fraction of cost, number or popularity of projects that are exclusively winning \emph{within} the impact area $l$ with respect to the cost, number of popularity of winning projects in that impact area:

\begin{table}[H]
\captionsetup{labelformat=empty}
\label{w}
\centering
\begin{tabular}{lll}

Cost Within Novelty  & Projects Within Novelty  & Popularity Within Novelty \\
\midrule
${S}_{o,c,l}^{\omega, f}$  = $\frac{\sum_{p=1}^{|{\widehat{W}_{l,f}}|} {c}_{p}}{\sum_{p=1}^{|{W_{l,f}}|} {c}_{p}}$ & ${S}_{o,n,l}^{\omega,f}$  = $\frac{|{\widehat{W}_{l,f}}|}{|{W_{l,f}}|}$ & ${S}_{o,v,l}^{\omega, f}$  = $\frac{\sum_{p=1}^{|{\widehat{W}_{l,f}}|} {v}_{p}}{\sum_{p=1}^{|{W_{l,f}}|} {v}_{p}}$ \\
\end{tabular}
\end{table}

\noindent \textbf{{\em Within-novelty} metrics calculated at the voters' ballot level}: The novelty $\omega$ within  an impact area $l$ for an aggregation method $f$ at the ballot level is calculated as the fraction of cost or number of projects that are exclusively winning \emph{within} the impact area $l$ which is also approved by the voter $v$ with respect to the cost, number of popularity of all winning projects in that impact area:

\begin{table}[H]
\captionsetup{labelformat=empty}
\label{w}
\centering
\begin{tabular}{ll}

Cost Within Novelty  & Projects Within Novelty   \\
\midrule
${S}_{b,c,l}^{\omega,f}$  = $\frac{\sum_{p=1}^{|{\widehat{W}_{l,v,f}}|} {c}_{p}}{\sum_{p=1}^{|{W_{l,f}}|} {c}_{p}}$ & ${S}_{b,n,l}^{\omega,f}$  = $\frac{|{\widehat{W}_{l,v,f}}|}{|{W_{l,f}}|}$ \\
\end{tabular}
\end{table}

\noindent \textbf{{\em Between-novelty} metrics calculated at the voting outcome level}: The novelty $\beta$ between impact areas for an aggregation method $f$ is calculated as the fraction of cost, number or popularity of exclusively winning projects for an impact area $l$ with respect to the cost, number or popularity of all exclusively winning projects:

\begin{table}[H]
\captionsetup{labelformat=empty}
\label{w}
\centering
\begin{tabular}{lll}

Cost Between Novelty  & Projects Between Novelty  & Popularity Between Novelty \\
\midrule
${S}_{o,c,l}^{\beta,f}$  = $\frac{\sum_{p=1}^{|{\widehat{W}_{l,f}}|} {c}_{p}}{\sum_{p=1}^{|{\widehat{W}_{f}}|} {c}_{p}}$ & ${S}_{o,n,l}^{\beta,f}$  = $\frac{|{\widehat{W}_{l,f}}|}{|{\widehat{W}_{f}}|}$ & ${S}_{o,v,l}^{\beta,f}$  = $\frac{\sum_{p=1}^{|{\widehat{W}_{l,f}}|} {v}_{p}}{\sum_{p=1}^{|{\widehat{W}_{f}}|} {v}_{p}}$ \\
\end{tabular}
\end{table}

\noindent \textbf{{\em Between-novelty} metrics calculated at the voters' ballot level}: The novelty $\beta$ between impact areas for an aggregation method $f$ at the ballot level is calculated as the fraction of cost or number of projects that are exclusively winning for an impact area $l$ which is also approved by a voter $v$ with respect to the cost or number of all exclusively winning projects:

\begin{table}[H]
\captionsetup{labelformat=empty}
\label{w}
\centering
\begin{tabular}{ll}
Cost Between Novelty  & Projects Between Novelty  \\
\midrule
${S}_{b,c,l}^{\beta,f}$  = $\frac{\sum_{p=1}^{|{\widehat{W}_{l,v,f}}|} {c}_{p}}{\sum_{p=1}^{|{\widehat{W}_{f}}|} {c}_{p}}$ & ${S}_{b,n,l}^{\beta,f}$  = $\frac{|{\widehat{W}_{l,v,f}}|}{|{\widehat{W}_{f}}|}$ \\

\end{tabular}
\end{table}

% Table for impact loss calculation of each respective metric

\begin{table}[H]
\centering
\caption{Loss or gain calculation for impact and novelty metrics }
\label{impact_loss_calculation}
\resizebox{0.84\textwidth}{!}{
    \begin{tabular}{lllllll}
    \toprule
    \multirow{2}{*}{Impact} & \multicolumn{2}{c}{Impact Loss [Cost]}  & \multicolumn{2}{c}{Impact Loss [Projects]}& \multicolumn{2}{c}{ Impact Loss [Popularity]} \\
    \cmidrule(lr){2-7}
          & Outcome & Individual Ballot & Outcome & Individual Ballot & Outcome \\
    \midrule
    
    Share &	$\textrm{S}_\textrm{o,c,l}^\textrm{i,$u$} - \textrm{S}_\textrm{o,c,l}^\textrm{i,$e$}$ &	$\textrm{S}_\textrm{b,c,l}^\textrm{i,$u$} -  \textrm{S}_\textrm{b,c,l}^\textrm{i,$e$}$ &	$\textrm{S}_\textrm{o,n,l}^\textrm{i,$u$} - \textrm{S}_\textrm{o,n,l}^\textrm{i,$e$}$ 	&	$\textrm{S}_\textrm{b,n,l}^\textrm{i,$u$} -  \textrm{S}_\textrm{b,n,l}^\textrm{i,$e$}$ & $\textrm{S}_\textrm{o,v,l}^\textrm{i,$u$} - \textrm{S}_\textrm{o,v,l}^\textrm{i,$e$}$ \\ 
    Representation &	$\textrm{R}_\textrm{o,c,l}^\textrm{i,$u$} - \textrm{R}_\textrm{o,c,l}^\textrm{i,$e$}$ &	$\textrm{R}_\textrm{b,c,l}^\textrm{i,$u$} -  \textrm{R}_\textrm{b,c,l}^\textrm{i,$e$}$ &	$\textrm{R}_\textrm{o,n,l}^\textrm{i,$u$} - \textrm{R}_\textrm{o,n,l}^\textrm{i,$e$}$ 	&	$\textrm{R}_\textrm{b,n,l}^\textrm{i,$u$} -  \textrm{R}_\textrm{b,n,l}^\textrm{i,$e$}$ & $\textrm{R}_\textrm{o,v,l}^\textrm{i,$u$} - \textrm{R}_\textrm{o,v,l}^\textrm{i,$e$}$ \\  
    Proportionality &	$\textrm{P}_\textrm{o,c,l}^\textrm{i,$u$} - \textrm{P}_\textrm{o,c,l}^\textrm{i,$e$}$ &	$\textrm{P}_\textrm{b,c,l}^\textrm{i,$u$} -  \textrm{P}_\textrm{b,c,l}^\textrm{i,$e$}$ &	$\textrm{P}_\textrm{o,n,l}^\textrm{i,$u$} - \textrm{P}_\textrm{o,n,l}^\textrm{i,$e$}$ 	&	$\textrm{P}_\textrm{b,n,l}^\textrm{i,$u$} -  \textrm{P}_\textrm{b,n,l}^\textrm{i,$e$}$ & $\textrm{P}_\textrm{o,v,l}^\textrm{i,$u$} - \textrm{P}_\textrm{o,v,l}^\textrm{i,$e$}$ \\  
    \midrule
    
    \multirow{2}{*}{Novelty } & \multicolumn{2}{c}{Novelty Loss [Cost]}  & \multicolumn{2}{c}{Novelty Loss [Projects]}& \multicolumn{2}{c}{Novelty Loss [Popularity]} \\
    \cmidrule(lr){2-7}
          & \multicolumn{2}{c}{Outcome}  & \multicolumn{2}{c}{Outcome} & Outcome \\
    \midrule
    
    Between-Novelty & \multicolumn{2}{c}{$\textrm{S}_\textrm{o,c,l}^\textrm{$\beta$,$u$} - \textrm{S}_\textrm{o,c,l}^\textrm{$\beta$,$e$}$} & \multicolumn{2}{c}{$\textrm{S}_\textrm{o,n,l}^\textrm{$\beta$,$u$} - \textrm{S}_\textrm{o,n,l}^\textrm{$\beta$,$e$}$} & $\textrm{S}_\textrm{o,v,l}^\textrm{$\beta$,$u$} - \textrm{S}_\textrm{o,v,l}^\textrm{$\beta$,$e$}$ \\

    Within-Novelty & \multicolumn{2}{c}{$\textrm{S}_\textrm{o,c,l}^\textrm{$\omega$,$u$} - \textrm{S}_\textrm{o,c,l}^\textrm{$\omega$,$e$}$} & \multicolumn{2}{c}{$\textrm{S}_\textrm{o,n,l}^\textrm{$\omega$,$u$} - \textrm{S}_\textrm{o,n,l}^\textrm{$\omega$,$e$}$} & $\textrm{S}_\textrm{o,v,l}^\textrm{$\omega$,$u$} - \textrm{S}_\textrm{o,v,l}^\textrm{$\omega$,$e$}$ \\
      
    \bottomrule
    \end{tabular}
}
\end{table}

\subsection{Threats to Validity}\label{sec:threats-to-validity}
The 345 real-world participatory budgeting instances are retrieved from the Pabulib repository and cover voting scenarios at city, district, and municipal level. These instances feature a broad spectrum of projects, from 5 to 220. The number of voters spans from 200 to 40,000, while the available budget expands in the range of 10K to 900K in PLN (Polish Zlotych). This diversity provides strong empirical foundation to scale up and generalize the insights derived from the impact and novelty analysis.

To account for the different nature of the voting instances, the measurements illustrated in the figures of this paper show the distribution over all voting instances, rather than only the mean values. Moreover, each proposed project in a given election instance is independently labelled based on the quartile it belongs within that particular instance irrespective of the cost of other proposed projects in different election instances. This normalization allows us to make comparisons among different voting instances that may come with costs of different levels.

For the participatory budgeting instances of (i) City Idea, Aarau and (ii) Green Million, Wieliczka, the impact area labels for the proposed projects were absent. As such, the impact areas labels for the proposed projects in these instances were assigned independently by multiple individuals to cross-validate the classification.

\section*{Acknowledgements}{The authors would like to thank the reviewers of this manuscript for their invaluable feedback. This work is funded by a UKRI Future Leaders Fellowship (MR\-/W009560\-/1): \emph{Digitally Assisted Collective Governance of Smart City Commons--ARTIO}'. All data used for this study are in public domain.}

\section*{Data and Code Availability}
Relevant dataset and source code used for the analysis used in this paper is made available at \noindent {\url{https://github.com/TDI-Lab/Impact-Analysis-Participatory-Budgeting}. 
}

\bibliographystyle{unsrt}
\bibliography{sample}

\makeatletter\@input{yy.tex}\makeatother
\end{document}

\maketitle
\tableofcontents
\section{Impact assessment by controlling cost and popularity}

\begin{figure}[!htb]
\centering\includegraphics[width=0.9\textwidth]{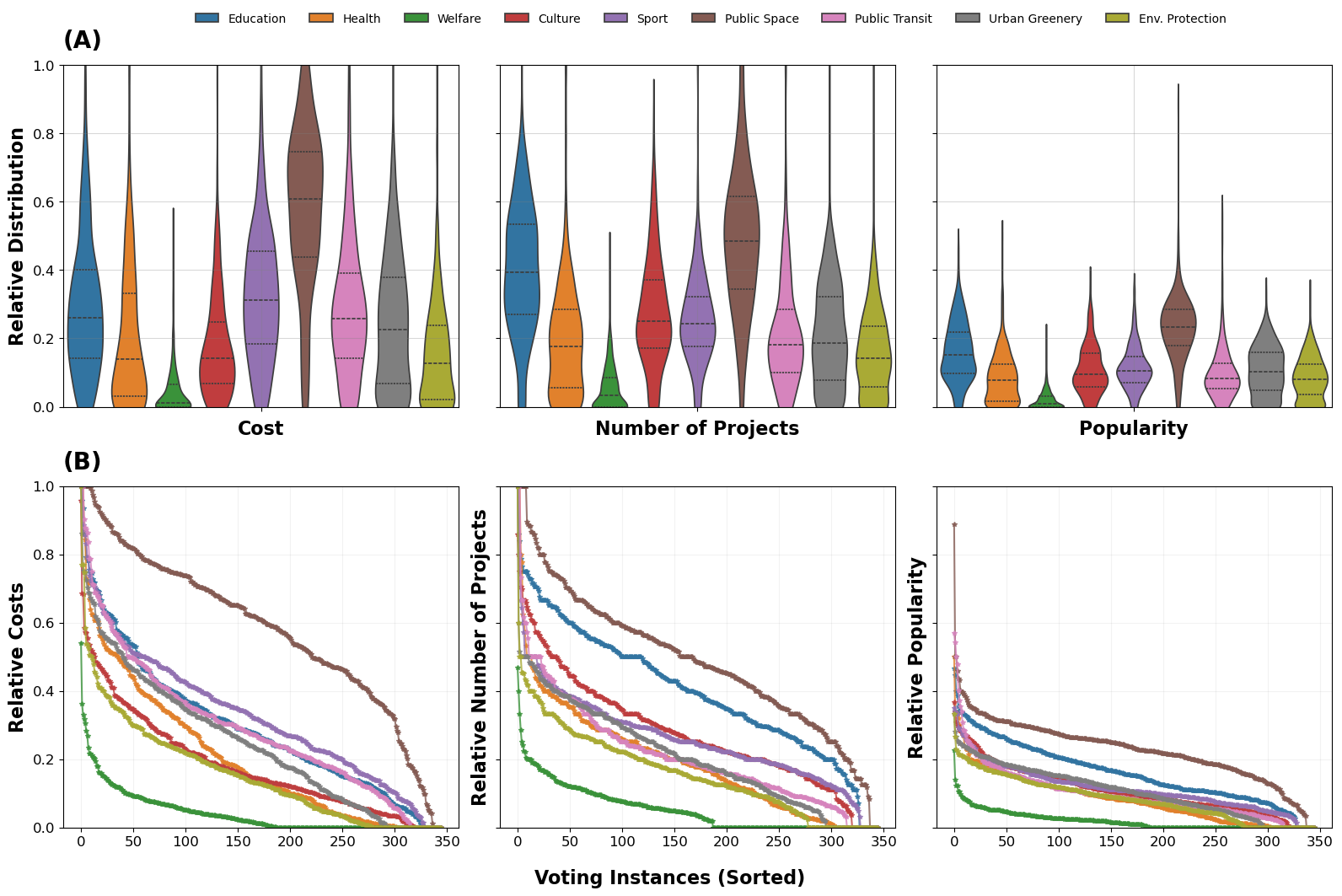}
\caption{\textbf{Distribution of projects, costs and their popularity among voters across different impact areas for 345 Pabulib election data.} The figure illustrates (A) the overall share of project costs, number of projects and their popularity among voters across different impact areas. (B) the sorted fractions of project cost, number of projects and their popularity for each impact area across the observed instances.}
\label{project_distribution}
\end{figure}

\noindent \textbf{Uneven distribution of number of projects, costs and popularity across different impact areas.} Figure~\ref{project_distribution} illustrates the distribution of projects, their cost and popularity among voters across different impact areas. The public space projects are costlier, more frequent as well as more popular in comparison with other impact areas. Education projects rank second in terms of cost, frequency as well as popularity among voters. Sport, public transit, urban greenery, health, environmental protection and culture projects follow next, while welfare projects are the least frequent, least popular and among the cheapest overall.
\\

\begin{figure}[!htb]
\centering\includegraphics[width=0.9\textwidth]{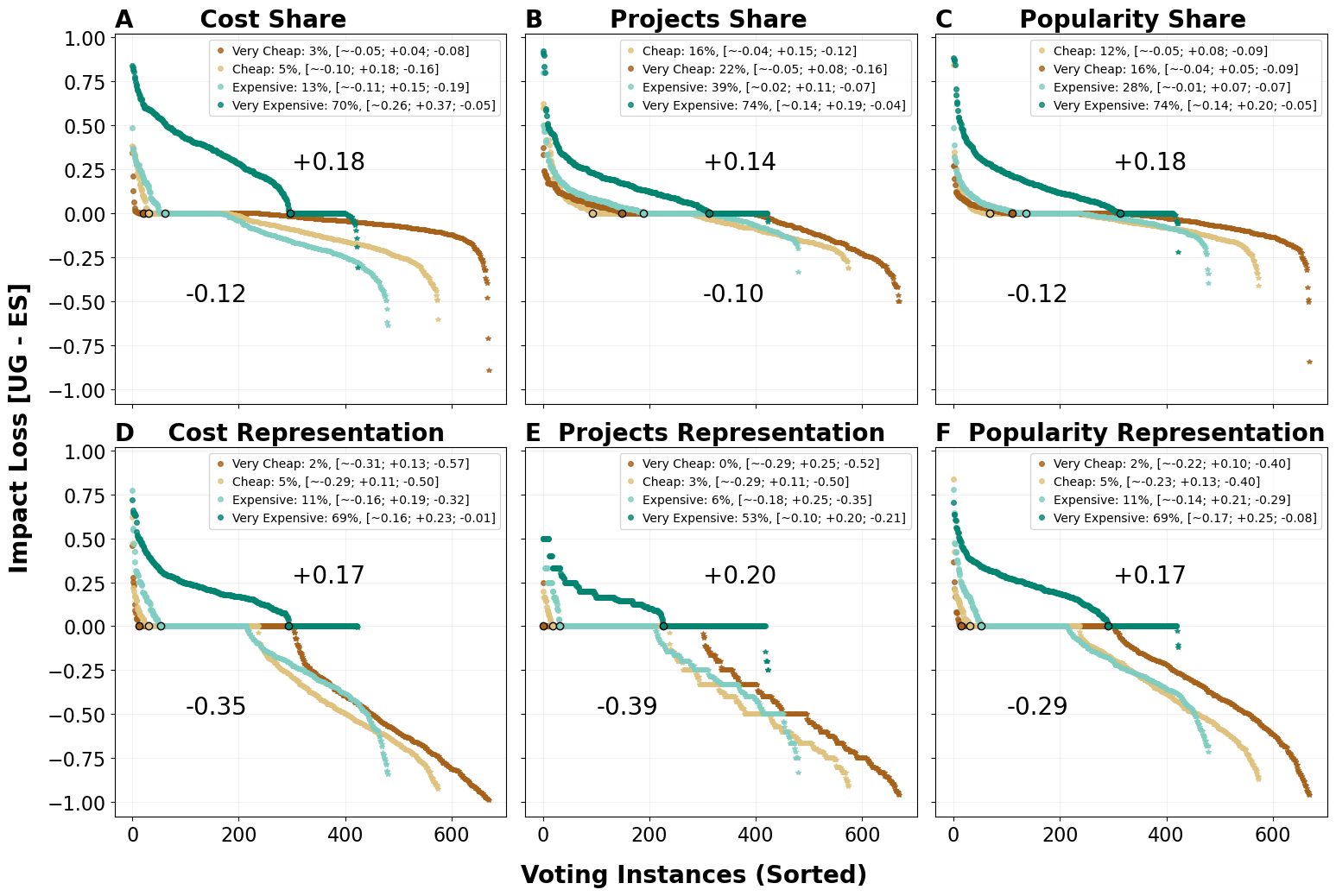}
\caption{\textbf{Equal shares results in voting outcomes with an impact loss in very expensive projects for share and representation metrics. Impact gain is more frequent and larger in scale for representation metrics, while impact loss is larger in scale for share metrics.} Share and representation metrics in terms of cost, number and popularity of projects are depicted as -- (A) cost share, (B) projects share, (C) popularity share, (D) cost representation (E) projects representation and (F) popularity representation. Positive loss (Y-axis, \textit{UG - ES}) for a cost quartile (colored lines) denotes better share and representation of such quartiles by utilitarian greedy (UG), while negative loss denotes better share and representation by equal shares (ES). The X-axis denotes voting instances sorted according to impact loss. A circular marker is placed along each line to easier distinguish the number of voting instances with positive/negative loss. The four numbers next to each cost level denote the (i) \% of voting instances with positive loss, (ii) the mean `$\mathtt{\sim}$', (iii) mean positive `+' and (iv) mean negative `-' impact loss. Two additional numbers with the prefixes '+' and '-' placed on each of the metrics signify the overall positive and negative impact loss respectively across all project cost quartiles.}
\label{project_cost_base_metrics}
\end{figure}

\noindent \textbf{Equal shares results in a dominant selection of very cheap and inexpensive projects but discourages the selection of very expensive projects.} Figure~\ref{project_cost_base_metrics} illustrates the impact loss or gain by equal shares for projects grouped on the basis of project cost levels in terms of the proposed impact metrics. The impact loss by equal shares method for very expensive projects is greater in magnitude compared to the impact gain for very cheap, cheap and slightly expensive projects in terms of cost share and projects share. For instance, the impact loss for very cheap, cheap and slightly expensive projects by equal shares is -5\%, -10\%, -11\% and -4\%, -5\%, 2\% for cost share and projects share respectively. The voting instances with cost share losses and projects share losses for these cost-quartile projects are only as few as 3\%, 5\%, 12\% and 16\%, 22\%, 39\% respectively. On the other hand, the cost share losses and projects share losses for very expensive projects is 26\% and 14\% respectively. Such losses for cost share and projects share is observed in 70\% and 74\% of the election instances respectively. \\

\begin{figure}[!htb]
\centering\includegraphics[width=0.9\textwidth]{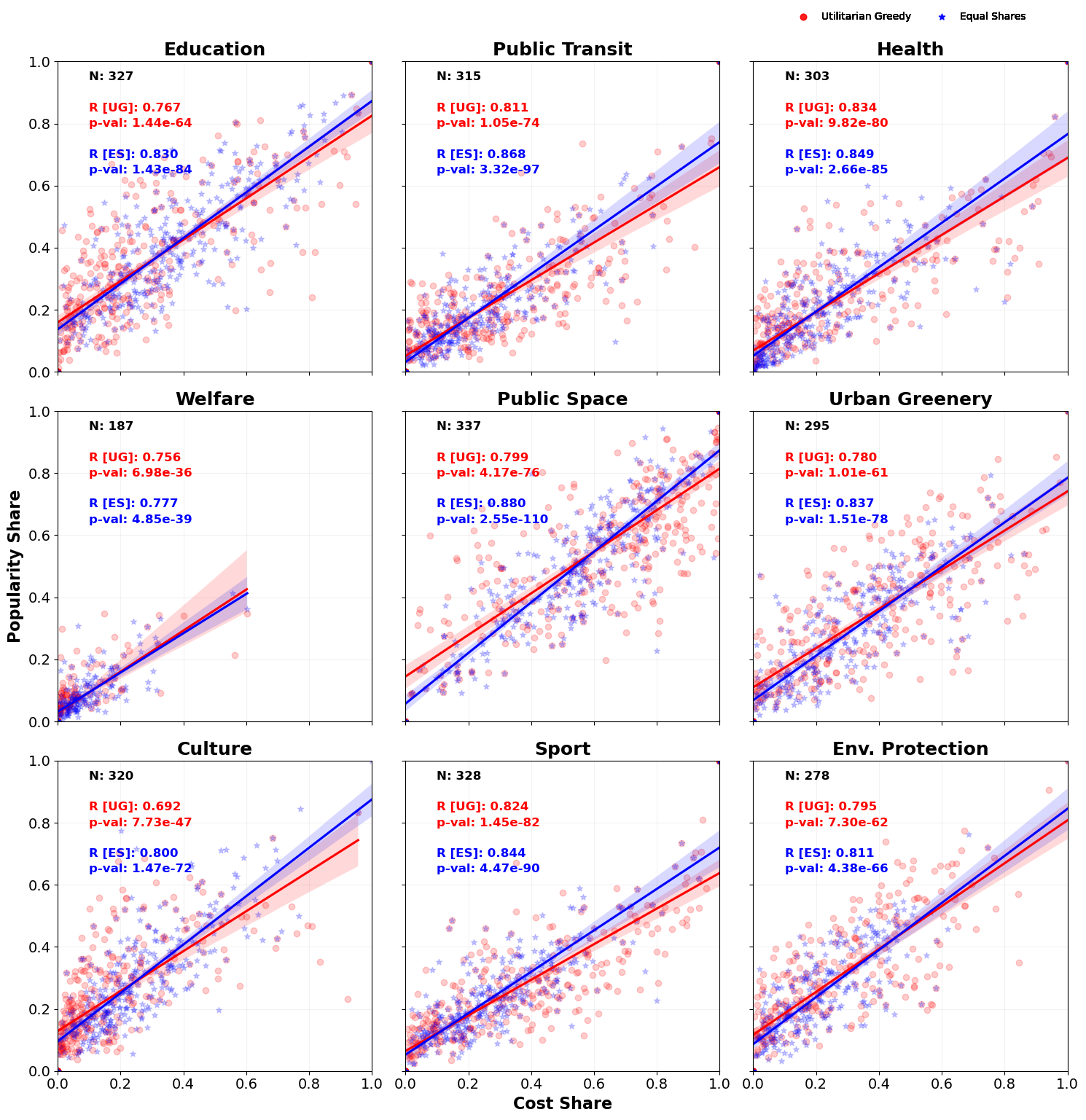}
\caption{\textbf{Winning outcomes by utilitarian greedy exhibit higher correlation between cost share and popularity share for projects related to public transit, public space, environmental protection and health related projects, while welfare and cultural projects show relatively lower degrees of correlation. On the other hand, equal shares shows higher correlation measure between cost share and popularity share across all impact areas.} For each impact area, a red circular marker (utilitarian greedy) and blue asterisk marker (equal shares) denotes cost share (X-axis) and the corresponding popularity share (Y-axis) of winning projects in a voting instance. Likewise, $N$ denotes the sample size for the voting instance with corresponding impact areas. For each of equal shares and utilitarian greedy, $R$ denotes the Pearson's correlation coefficient between cost share and popularity share and corresponding p-values are shown.}
\label{cost_share_vs_popularity_share}
\end{figure}

\noindent \textbf{Correlation between cost and popularity of projects across different impact areas.} Figure~\ref{cost_share_vs_popularity_share} shows the correlation measure between the cost share and popularity share of winning projects across different impact areas for the observed voting instances. Projects related to sustainable infrastructure such as public transit, public space and urban greenery exhibit a higher correlation coefficient, while culture, welfare and education projects show relatively lower degrees of correlation. On account of equal distribution of cost with respect to votes (which amounts to overall popularity), the winning outcomes in equal shares display higher degrees of correlation across all impact areas and specifically improving on the culture and welfare projects.

\begin{figure}[!htb]
\centering\includegraphics[width=\textwidth]{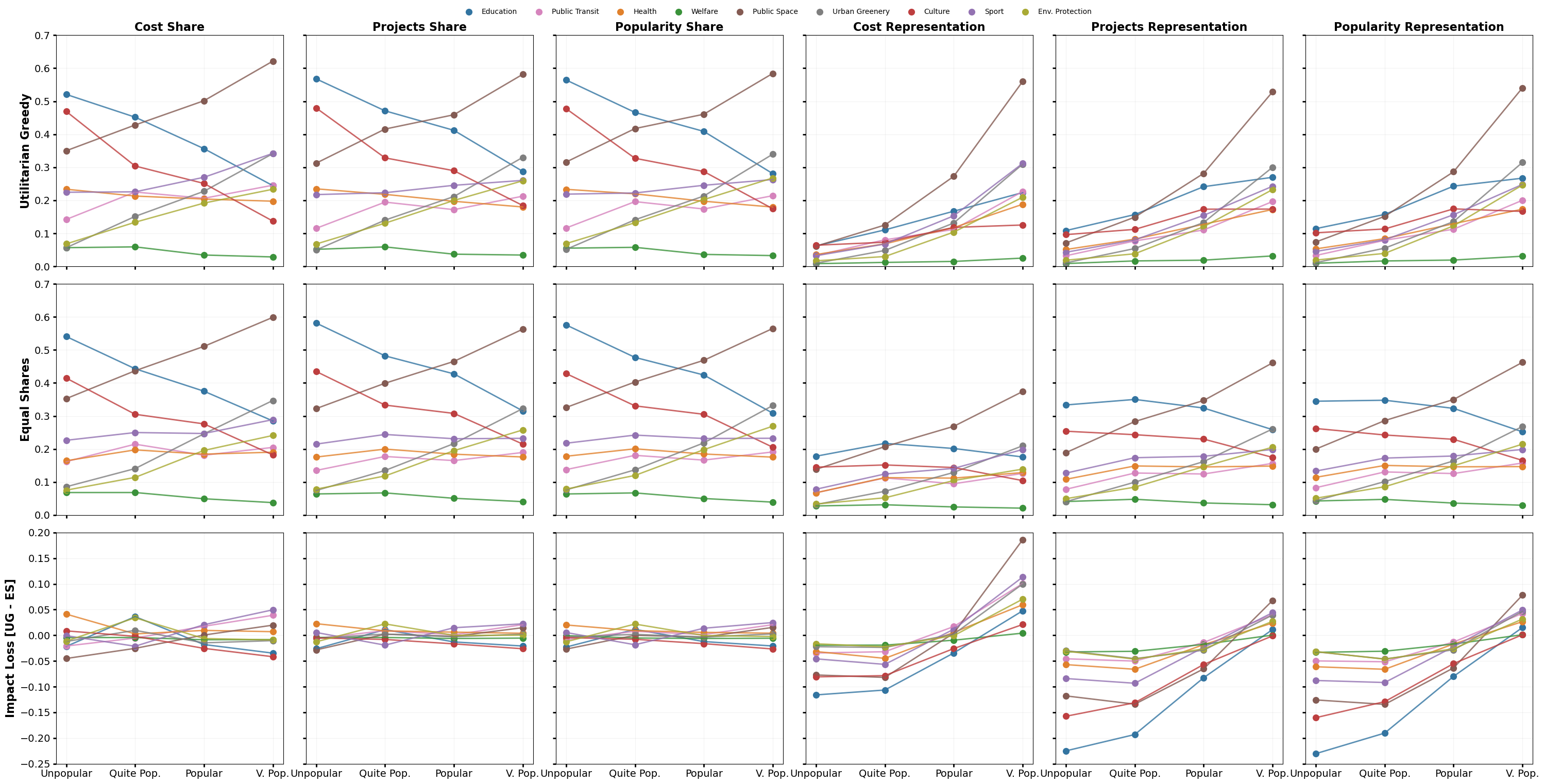}
\caption{\textbf{
Impact loss with equal shares is significant for very popular public space projects. Impact gain by equal shares for welfare projects is similar across all popularity levels.} Metrics that capture the winning set i.e. cost share, projects share and popularity share do not exhibit significant gain or losses, however metrics that deal with proposal set i.e., cost representation, projects representation and popularity representation show significant gain and losses for unpopular and very popular projects respectively. Impact performance values (Y-axis) of
equal shares (ES) and utilitarian greedy (UG) across different impact areas for (A) cost share, (B) projects share, (C) popularity share, (D) cost representation, (E) projects representation and (F) popularity representation at different popularity levels (X-axis) is shown. The last column depicts the respective impact loss values \emph{(UG-ES)}.}
\label{popularity_control_impact_performance}
\end{figure}

\noindent \\ \textbf{Impact loss or gain across different impact areas at different levels of projects popularity.} For each independent voting instance, all proposed projects were grouped into four separate levels of popularity -- \emph{unpopular, quite popular, popular and very popular} based on quartile distribution of votes received by each project in that instance. Figure~\ref{popularity_control_impact_performance} illustrates the performance of equal shares and utilitarian greedy in terms of the proposed impact metrics when winning set of projects are evaluated across different impact areas at different levels of popularity. For the impact metrics of cost share and projects share (which captures the winning sets), the performance of equal shares and utilitarian greedy for a specific impact area at a specific level of popularity remains similar. This is evident from the impact loss column (3rd column) having the range of values between -0.4 to 0.4 for cost share and projects share. However for public transit and public space projects, there is an upward trend from impact gain (negative impact loss) to impact loss (positive) as projects become more popular. The opposite holds for cultural projects. In terms of cost representation and projects representation (which deals with proposal sets), there is significant impact loss for very popular projects while there is impact gain for unpopular projects across all impact areas. An impact loss of upto 19\%, 7\% for very popular projects and an impact gain of upto 12\%, 22\% for unpopular projects is observed in terms of cost representation and projects representation respectively. The highest impact gain for unpopular projects belongs to impact areas of education and culture, while the highest impact loss for very popular projects is observed in public space. 

\clearpage
\section{Impact loss or gain measured at the voting outcome level}\label{sec:study}
\begin{figure}[!htb]
\centering\includegraphics[width=0.7\textwidth]{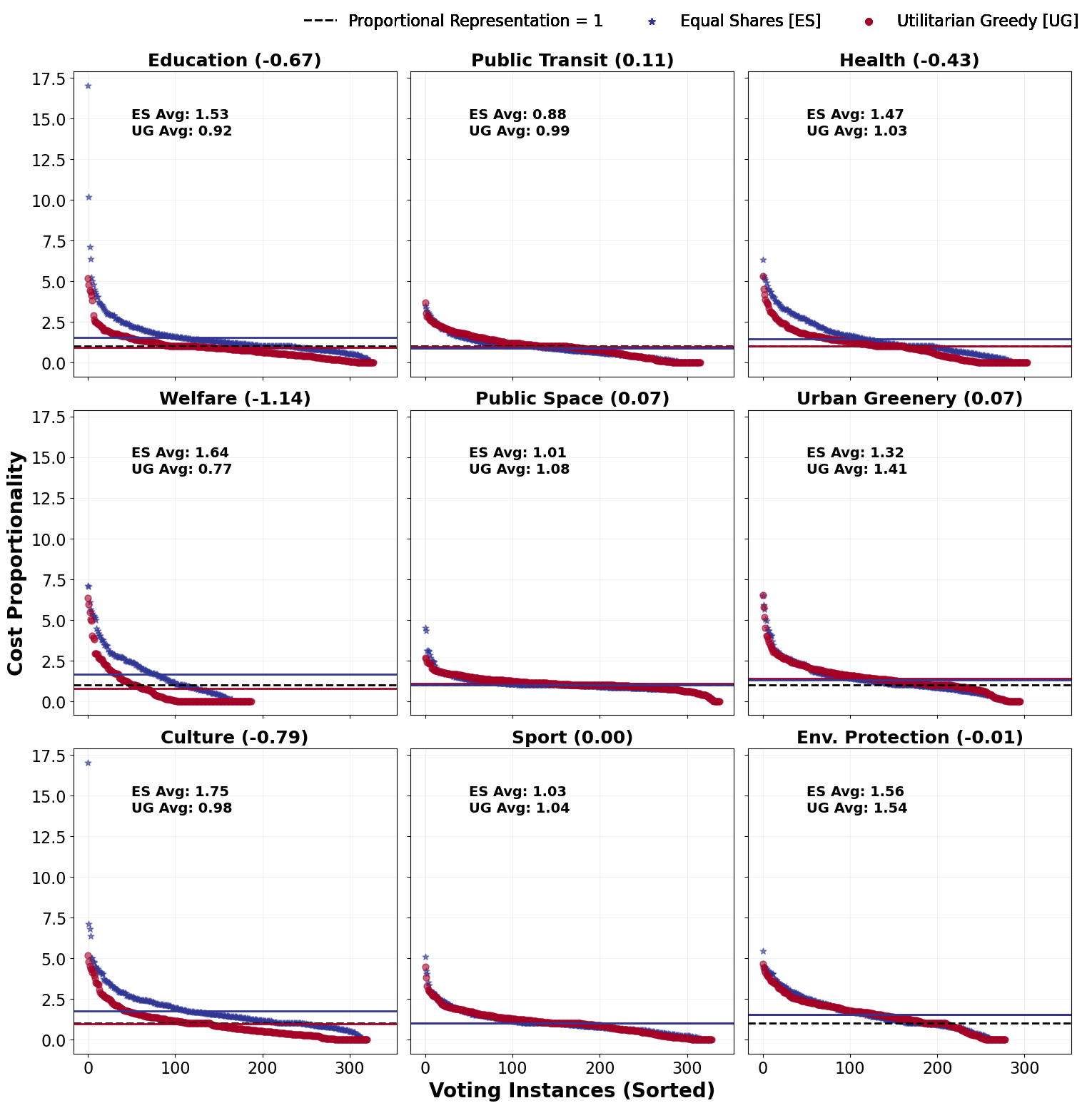}
\caption{\textbf{
Equal shares (ES) shows impact gain in cost proportionality for education, welfare, culture and health, with an impact loss in all other impact areas.} For each of the impact areas, the X-axis denotes voting instances sorted according to the values of cost proportionality (Y-axis). A horizontal dotted black line at y=1 denotes a proportional cost representation. Average values for cost proportionality by utilitarian greedy and equal shares is shown for each impact area. The number in parentheses next to each impact area denotes the relative loss in cost proportionality by equal shares compared to utilitarian greedy. i.e. \emph{(UG-ES) / UG}.}
\label{cost_proportionality}
\end{figure}

\begin{figure}[!htb]
\centering\includegraphics[width=0.7\textwidth]{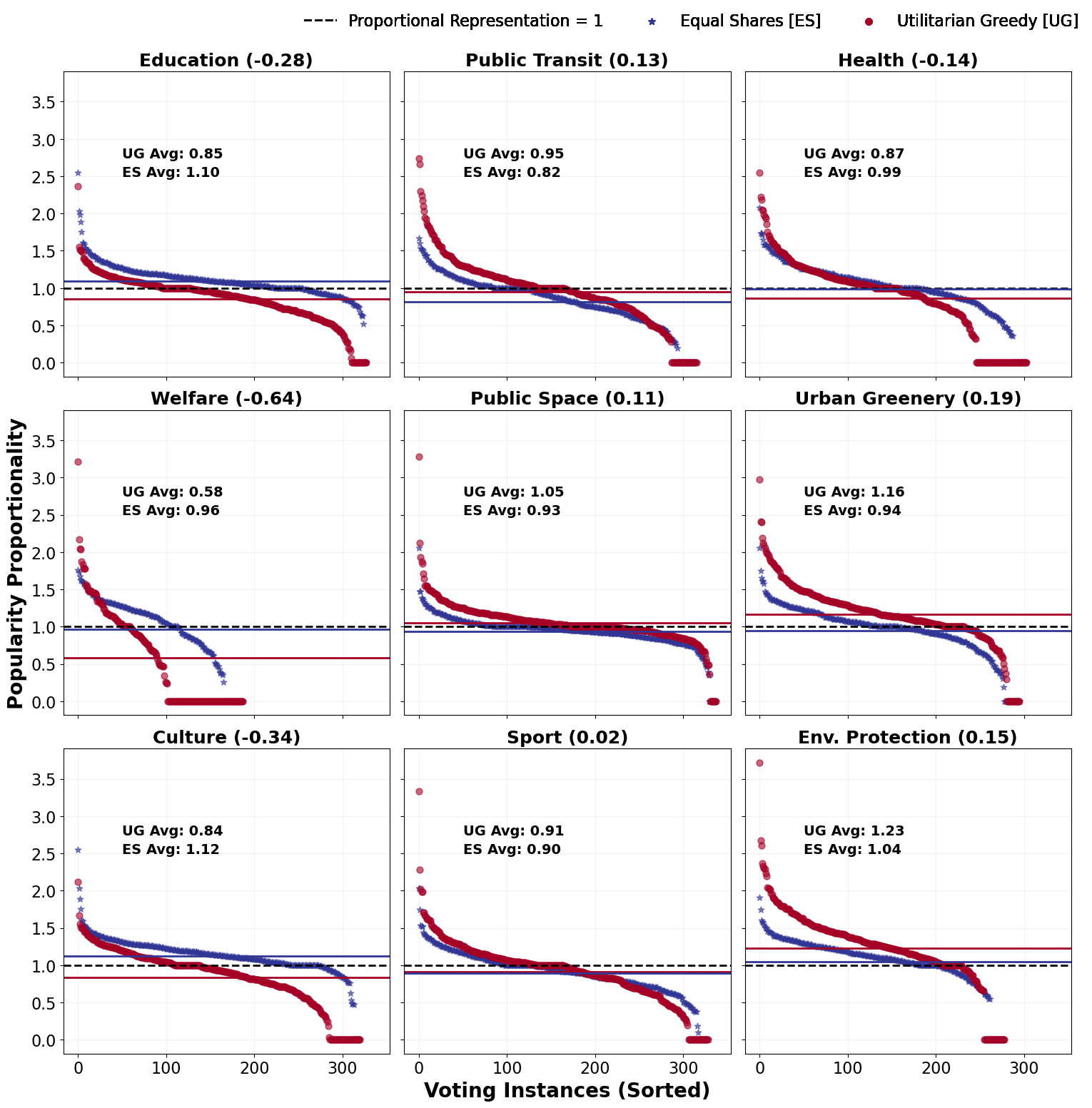}
\caption{\textbf{Equal shares (ES) shows impact gain in popularity proportionality for education, welfare, culture and health, with an impact loss in all other impact areas. The popularity proportionality metric follows similar trends with respect to cost proportionality metrics such that impact gain is observed for education, culture and welfare projects while an impact loss is observed for public transit projects.} For each of the impact areas, the X-axis denotes voting instances sorted according to the values of popularity proportionality (Y-axis). A horizontal dotted black line at y=1 denotes proportional popularity representation. Average values for popularity proportionality by utilitarian greedy and equal shares are shown for each impact area. The number in parentheses next to each impact area denotes the relative loss in popularity proportionality by equal shares compared to utilitarian greedy. i.e. \emph{(UG-ES) / UG}.}
\label{popularity_proportionality}
\end{figure}

\noindent \textbf{Equal shares results in a proportional representation of popularity across different impact areas in the winning set compared to utilitarian greedy. However, winning cost by equal shares for education, culture and welfare is over-represented but for public transit projects is under-represented.} Figure~\ref{cost_proportionality} shows the cost proportionality while Figure~\ref{popularity_proportionality} shows the popularity proportionality of the winning outcomes by equal shares and utilitarian greedy across different impact areas. For cost proportionality, the winning costs by equal shares are over-represented for education, culture and health, while cost is under-represented for public transit projects. For popularity proportionality, it is observed that across all impact areas equal shares results in a more proportional representation of votes compared to utilitarian greedy. \emph{For public transit projects, there is a significant under-representation of projects in the winning set by equal shares when compared with the popularity of such projects.} \\

\clearpage
\section{Impact loss or gain measured at the voters' ballot level}

\begin{figure}[!htb]
\centering\includegraphics[width=0.75\textwidth]{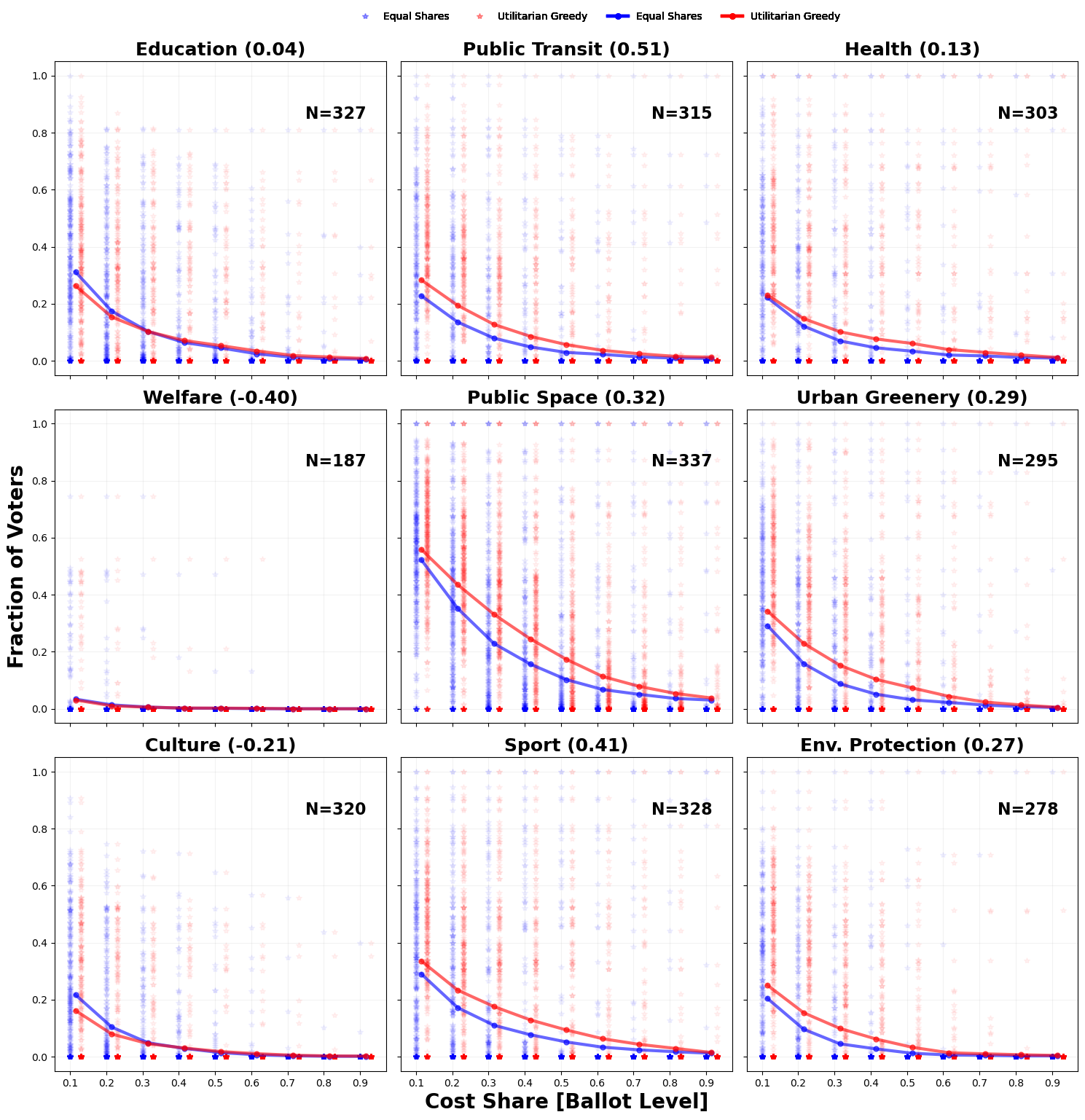}
\caption{\textbf{ Equal shares preserves the voters’ cost share for projects with impact gain,
while the cost share is reduced for projects with impact loss.} For each impact area, the fraction of voters (Y-axis) for different levels of projects cost share (X-axis) is shown for utilitarian greedy and equal shares. A total of 345 approval voting instances are counted, with the value of $N$ denoting the number of election instances with at least one proposed project belonging to the corresponding impact area. The numbers in the parentheses next to each impact area denotes the relative loss measured by the mean difference of cost share at the ballot level between utilitarian greedy and equal shares with respect to the value of utilitarian greedy i.e. \emph{(UG-ES) / UG}.}
\label{ballot_share_cost}
\end{figure}

\begin{figure}[!htb]
\centering\includegraphics[width=0.8\textwidth]{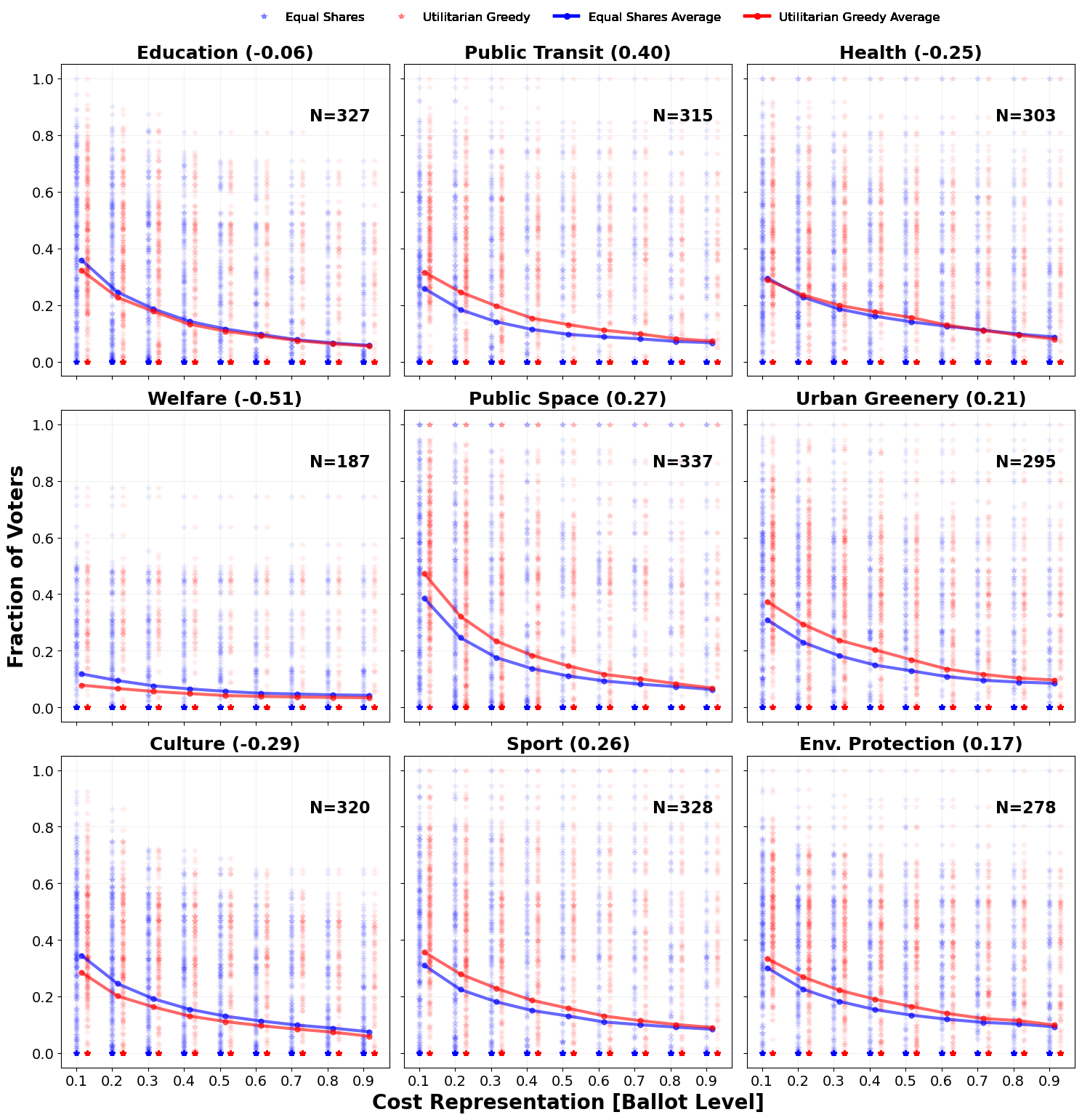}
\caption{\textbf{ Equal shares shows increased voters' cost representation for projects with impact gain, while the cost representation is reduced for projects with impact loss.} For each impact area, the fraction of voters (Y-axis) for different levels of projects cost representation (X-axis) is shown for utilitarian greedy and equal shares. A total of 345 approval voting instances are counted, with the value of $N$ denoting the number of election instances with at least one proposed project belonging to the corresponding impact area. The numbers in the parentheses next to each impact area denote the relative loss measured by the mean difference of cost representation at the ballot level between utilitarian greedy and equal shares with respect to the value of utilitarian greedy i.e., \emph{(UG-ES) / UG}.}
\label{ballot_representation_cost}
\end{figure}

\begin{figure}[!htb]
\centering\includegraphics[width=0.9\textwidth]{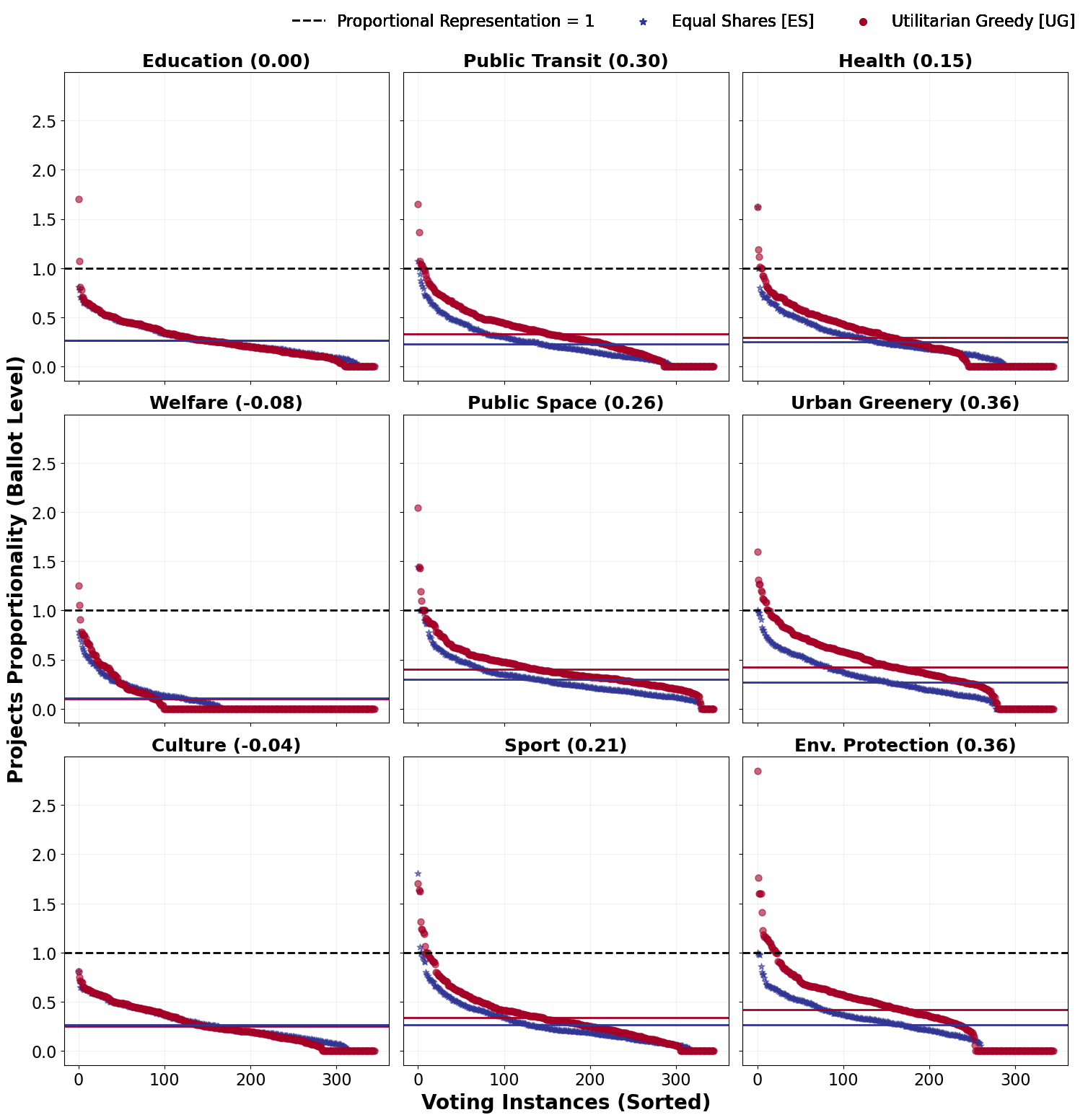}
\caption{ \textbf{ Equal shares preserves the voters’ projects proportionality at the ballot level with impact gain for impact areas such as education and culture, while the projects proportionality is reduced under projects with impact loss.} For each of the impact areas, the X-axis denotes voting instances sorted according to the values voter proportionality (Y-axis). A horizontal dotted black line at y=1 denotes proportional projects representation. Average values for projects proportionality by utilitarian greedy and equal shares is shown as horizontal lines across each impact area. The number in parentheses next to each impact area denotes the relative loss in projects proportionality by equal shares compared to utilitarian greedy. i.e. \emph{(UG-ES) / UG}.}
\label{ballot_proportionality_project}
\end{figure}

\begin{figure}[!htb]
\centering\includegraphics[width=0.9\textwidth]{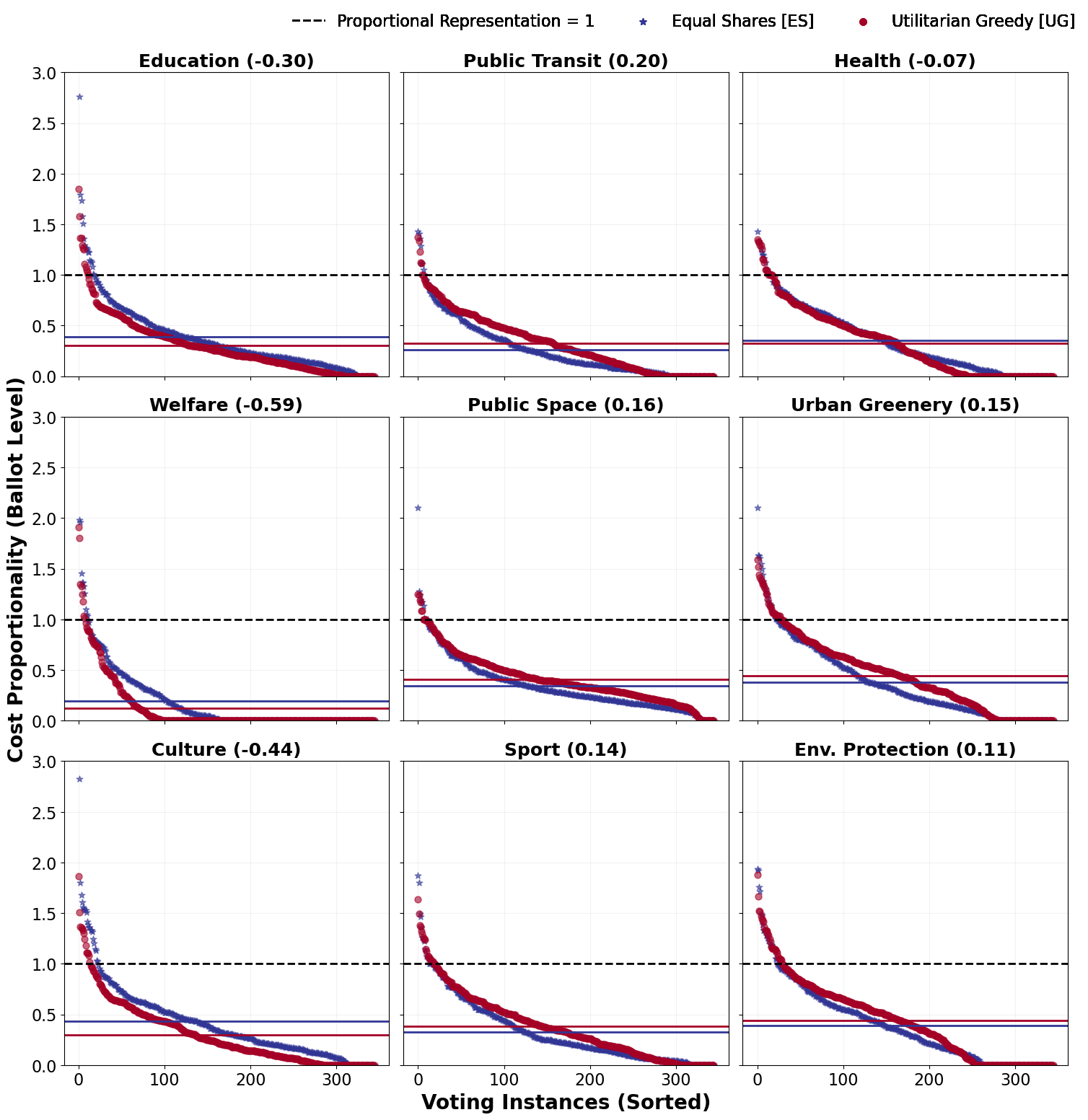}
\caption{ \textbf{ Equal shares preserves the voters’ cost proportionality with impact gain, while
the cost proportionality is reduced under projects with impact loss} For each of the impact areas, the X-axis denotes voting instances sorted according to the values of cost proportionality at the ballot level (Y-axis). A horizontal dotted black line at y=1 denotes proportional cost representation. Average values for cost proportionality by utilitarian greedy and equal shares is shown as horizontal lines across each impact area. The number in parentheses next to each impact area denotes the relative loss in cost proportionality by equal shares compared to utilitarian greedy. i.e. \emph{(UG-ES) / UG}.}
\label{ballot_proportionality_cost}
\end{figure}

This section presents the findings from the analysis of the impact gain or loss calculated at individual voters' ballot level across different impact areas. These metrics are used to understand how voters' choice relate to the voting outcomes and how impact gain or loss aligns at the individual voters' ballot and the voting outcome level. Figure~\ref{ballot_share_cost} shows cost share and Figure~\ref{ballot_representation_cost} shows cost representation by equal shares and utilitarian greedy across different impact areas at the individual voters' ballot level. Also, Figure~\ref{ballot_proportionality_project} show projects proportionality and Figure~\ref{ballot_proportionality_cost} shows cost proportionality across different impact areas by equal shares and utilitarian greedy at the individual voters' ballot level.

\clearpage
\section{Impact assessment in an exemplary voting scenario}

\noindent \\ An exemplary participatory budgeting voting scenario is illustrated in Table~\ref{toy_voting_example}  with 11 voters, 5 projects with impact area labels and a budget  of \$1000.  The proposed metrics are calculated in Table~\ref{metric_calculation_example}.

% Tabular data for voting example
\begin{table}[H]
% \captionsetup{justification=raggedright, singlelinecheck=false}
\caption{An exemplary participatory budgeting voting scenario for different projects with impact area labels}
\label{toy_voting_example}
\resizebox{1\textwidth}{!}{
\begin{tabular}{cccccccccccccccccc}
\toprule
\multirow{2}{*}{Projects} & \multirow{2}{*}{Impact Areas} & \multirow{2}{*}{Cost} & \multirow{2}{*}{Votes} & \multicolumn{11}{c}{Voters} 
 & \multirow{2}{*}{UG} & \multirow{2}{*}{ES} \\
\cmidrule(lr){5-15}
      & & &  & 1& 2& 3& 4& 5& 6& 7& 8& 9& 10& 11& & \\
\midrule
A& Education & 700& 7 &\checkmark& \checkmark& \checkmark& \checkmark& \checkmark& \checkmark& & & & & \checkmark& Y& Y \\
B& Welfare, Health & 400 & 6 & \checkmark& \checkmark& \checkmark& \checkmark& \checkmark& \checkmark& & & & & & Y& N \\
C&  Health& 250 & 5 & & \checkmark& & \checkmark& \checkmark& & \checkmark& & & \checkmark& & N& N \\
D& Education, Health& 200 & 4 & & & & & & & \checkmark& \checkmark& \checkmark& \checkmark& & N& Y \\
E& Education, Welfare& 100 & 3 & & & & & & & \checkmark& & \checkmark& \checkmark& & N& Y\\
\bottomrule
\end{tabular}}
\end{table}

\begin{table}[H]
% \captionsetup{justification=raggedright, singlelinecheck=false}
\caption{Impact and novelty metrics calculated at the voting outcome level for the voting scenario in Table~\ref{toy_voting_example}}
\label{metric_calculation_example}
\resizebox{1\textwidth}{!}{
\begin{tabular}{cccccccc}
\toprule
\multirow{2}{*}{Metric} & \multirow{2}{*}{Dimension} &  \multicolumn{2}{c}{Education}  & \multicolumn{2}{c}{Welfare} & \multicolumn{2}{c}{Health} \\
\cmidrule(lr){3-8}
& & UG & ES & UG & ES & UG & ES\\
\midrule
\multicolumn{8}{c}{Impact metrics at the voting outcome level}\\ \midrule \midrule
\multirow{3}{*}{Share} & Cost &	700/1100 &		1000/1000 &		400/1100 &		100/1000 &		400/1100 &		200/1000 \\
& Number of Projects	 &	1/2 &		3/3 &		1/2	 &	1/3 &		1/2	 &	1/3\\
& Popularity	& 7/13		& 14/14	& 	6/13	& 	3/14	& 	6/13		& 4/14 \\
\midrule
\multirow{3}{*}{Representation} & Cost &	700/1000 &		1000/1000	 &	400/500	 &	100/500	 &	400/850	 &	200/850 \\
& Number of Projects	 &	1/3		 &	3/3		 &	1/2		 &	1/2		 &	1/3	 &		1/3\\
& Popularity	& 7/14	 &	14/14	 &	6/9	 &	3/9 &		6/15 &		4/15 \\ 
\midrule
\multirow{3}{*}{Proportionality} & Cost &	(700/1100)/(1000/1650)	 &	(1000/1000)/(1000/1650) &		(400/1100)/(500/1650)	 &	(100/1000)/(500/1650) &		(400/1100)/(850/1650)	 &	(200/1000)/(850/1650) \\
& Number of Projects	 &	(1/2)/(3/5)  &	(3/3)/(3/5)	  &(1/2)/(2/5)	  &(1/3)/(2/5)	 & (1/2)/(3/5)  &	(1/3)/(3/5)\\
& Popularity	& (7/13)/(14/25)   &		(14/14)/(14/25)	   &	(6/13)/(9/25)	   &	(3/14)/(9/25)	  &	 (6/13)/(15/25)	   &	(4/14)/(15/25) \\ 
\midrule
\multicolumn{8}{c}{Impact metrics at the ballot level [Voter 1]}\\ \midrule \midrule

\multirow{2}{*}{Share} & Cost &	700/1100 &		700/1000 &		400/1100 &		0/1000 &		400/1100 &		0/1000 \\
& Number of Projects	 &	1/2	 &		1/3 &			1/2 &			0/3	 &		1/2 &			0/3 \\

\midrule

\multirow{2}{*}{Representation} & Cost &	700/1000	 &	700/1000 &		400/500	 &	0/500	 &	400/850	 &	0/850 \\
& Number of Projects	 &	1/3		 &	1/3	 &		1/2	 &		0/2	 &		1/3	 &		0/3 \\
\midrule

\multirow{2}{*}{Proportionality} & Cost &	(700/1100)/(1000/1650)	 &	(700/1100)/(1000/1650) 	 &	 (400/1100)/(500/1650)	 &		(0/1000)/(500/1650)		 &	(400/1100)/(850/1650)	 &		(0/1000)/(850/1650) \\
& Number of Projects	 &	(1/2)/(3/5) 	 &		(1/3)/(3/5) 	 &		(1/2)/(2/5)		 &	 (0/3)/(2/5)		 &	(1/2)/(3/5) 	 &		(0/3)/(3/5)\\
\midrule

\multicolumn{8}{c}{Novelty [{\em within impact areas}] metrics at the voting outcome level}\\ \midrule \midrule
\multirow{3}{*}{Share} & Cost &	0/700 &		300/1000 &		400/400	 &	100/100	 &	400/400	 &	200/200\\
& Number of Projects	 &	0/1	 &	2/3 &		1/1 &		1/1	 &	1/1	 &	1/1\\
& Popularity	& 0/7  &		7/14	 &	6/6	 &	3/3	 &	6/6  &		4/4 \\
\midrule

\multicolumn{8}{c}{Novelty [{\em between impact areas}] metrics at the voting outcome level}\\ \midrule \midrule
\multirow{3}{*}{Share} & Cost &	0/400	 &	300/300	 &	400/400	 &	100/300	 &	400/400	 &	200/300\\
& Number of Projects	 &	0/1	 &	2/2	 &	1/1	 &	1/2 &		1/1	 &	1/2\\
& Popularity	& 0/6	 &		7/7	 &		6/6		 &	3/7	 &		6/6		 &	4/7 \\
\midrule
\end{tabular}}
\end{table}

\section{Significance tests}
\noindent \\ Tables~\ref{paired_t_test_impact_metrics}, ~\ref{paired_t_test_proportionality_metrics} and ~\ref{paired_t_test_novelty_metrics} illustrate the paired t-test significance values for the outcome of the proposed metrics concerning the utilitarian greedy and equal shares. For 95\% of the cases, the outcome is significant and validates our observations and methodology. 
We also present significance values and relative importance values in  Table~\ref{conjoint_analysis_results} for the conjoint analysis to predict the budget utilization based on the impact areas.

% Tabular data for t-test results on impact metrics
\begin{table}[!htb]
\centering
\caption{Paired t-test statistics for impact assessments at the voting outcome level using share and representation metrics}
\label{paired_t_test_impact_metrics}
\begin{scriptsize}
\begin{tabular}{cccc}
\toprule
Impact Metrics & Impact Areas & t-statistic & p-value \\
\midrule
Cost Share & Education & -8.73 & 1.09 $\times$ 10 \textsuperscript{-16} \\
& Public Transit and Roads & 4.24 & 2.85 $\times$ 10 \textsuperscript{-5} \\
& Health & 1.12 & \textcolor{red}{0.26} \\
& Welfare & -5.45 & 9.58 $\times$ 10 \textsuperscript{-8} \\
& Public Space & 5.61 & 4.26 $\times$ 10 \textsuperscript{-8} \\
& Urban Greenery & 5.27 & 2.39 $\times$ 10 \textsuperscript{-7} \\
& Culture & -9.72 & 6.64 $\times$ 10 \textsuperscript{-20} \\
& Sport & 4.39 & 1.51 $\times$ 10 \textsuperscript{-5} \\
& Environmental Protection & 4.73 & 3.21 $\times$ 10 \textsuperscript{-6} \\
\midrule
Projects Share & Education & -12.19 & 1.17 $\times$ 10 \textsuperscript{-28} \\
& Public Transit and Roads & 4.38 & 1.59 $\times$ 10 \textsuperscript{-5} \\
& Health & 0.67 & \textcolor{red}{0.51} \\
& Welfare & -5.69 & 2.59 $\times$ 10 \textsuperscript{-8} \\
& Public Space & 8.70 & 1.35 $\times$ 10 \textsuperscript{-16} \\
& Urban Greenery & 9.59 & 1.74 $\times$ 10 \textsuperscript{-19} \\
& Culture & -11.31 & 1.99 $\times$ 10 \textsuperscript{-25} \\
& Sport & 1.48 & \textcolor{red}{0.14} \\
& Environmental Protection & 0.73 & 1.08 $\times$ 10 \textsuperscript{-16} \\
\midrule
Popularity Share & Education & -11.75 & 4.49 $\times$ 10 \textsuperscript{-27} \\
& Public Transit and Roads & 4.45 & 1.14 $\times$ 10 \textsuperscript{-5}\\
& Health & 0.71 & \textcolor{red}{0.47} \\
& Welfare & -5.94 & 6.61 $\times$ 10 \textsuperscript{-9}  \\
& Public Space & 7.07 & 8.89 $\times$ 10 \textsuperscript{-12} \\
& Urban Greenery & 7.75 & 1.02 $\times$ 10 \textsuperscript{-13} \\
& Culture & -11.59 & 1.76 $\times$ 10 \textsuperscript{-26} \\
& Sport & 2.43 & 0.0156 \\
& Environmental Protection & 7.29 & 2.15 $\times$ 10 \textsuperscript{-12}\\
\midrule
Cost Representation & Education & -9.33 & 1.32 $\times$ 10 \textsuperscript{-18} \\
& Public Transit and Roads & 3.26 & 0.001 \\
& Health & -4.74 & 3.20 $\times$ 10 \textsuperscript{-6} \\
& Welfare & -8.13 & 7.91 $\times$ 10 \textsuperscript{-15} \\
& Public Space & 4.24 & 2.92 $\times$ 10 \textsuperscript{-5} \\
& Urban Greenery & 3.43 & 0.0006 \\
& Culture & -11.92 & 1.11 $\times$ 10 \textsuperscript{-27} \\
& Sport & 1.16 & \textcolor{red}{0.25} \\
& Environmental Protection & 0.84 & \textcolor{red}{0.40} \\
\midrule
Projects Representation & Education & -19.57 & 6.59 $\times$ 10 \textsuperscript{-58} \\
& Public Transit and Roads & -5.32 & 1.91 $\times$ 10 \textsuperscript{-7} \\
& Health & -11.14 & 7.95 $\times$ 10 \textsuperscript{-25} \\
& Welfare & -11.07 & 1.42 $\times$ 10 \textsuperscript{-24} \\
& Public Space & -9.62 & 1.50 $\times$ 10 \textsuperscript{-19} \\
& Urban Greenery & -3.69 & 0.0002 \\
& Culture & -20.55 & 7.16 $\times$ 10 \textsuperscript{-62} \\
& Sport & -10.74 & 1.95 $\times$ 10 \textsuperscript{-23} \\
& Environmental Protection & -5.87 & 9.92 $\times$ 10 \textsuperscript{-9} \\
\midrule

Popularity Representation & Education & -16.69 & 2.65 $\times$ 10 \textsuperscript{-46}\\
& Public Transit and Roads & -2.04 & 0.0424 \\
& Health & -8.62 & 2.51 $\times$ 10 \textsuperscript{-16} \\
& Welfare & -10.18 & 1.89 $\times$ 10 \textsuperscript{-21}  \\
& Public Space & -4.15 & 4.23 $\times$ 10 \textsuperscript{-5} \\
& Urban Greenery & -0.22 & \textcolor{red}{0.82}  \\
& Culture & -17.75 & 1.54 $\times$ 10 \textsuperscript{-50} \\
& Sport & -6.31 & 8.36 $\times$ 10 \textsuperscript{-10}\\
& Environmental Protection & -2.93 & 0.0037 \\
\bottomrule
\end{tabular}
\end{scriptsize}
\end{table}

% Tabular data for t-test results on impact metrics
\begin{table}[!htb]
\centering
\caption{Paired t-test statistics for impact assessments at the voting outcome level using proportionality metrics. }
\label{paired_t_test_proportionality_metrics}
\begin{scriptsize}
\begin{tabular}{cccc}

\toprule
Impact Metrics & Impact Areas & t-statistic & p-value \\
\midrule
Cost Proportionality & Education & -8.13 & 9.06 $\times$ 10 \textsuperscript{-15} \\
& Public Transit and Roads & 2.15 & 0.0324 \\
& Health & -6.03 & 4.67 $\times$ 10 \textsuperscript{-9} \\
& Welfare & -8.39 & 1.14 $\times$ 10 \textsuperscript{-14}  \\
& Public Space & 2.06 & 0.0399 \\
& Urban Greenery & 1.51 & \textcolor{red}{0.1309}  \\
& Culture & -10.29 & 1.21 $\times$ 10 \textsuperscript{-21} \\
& Sport & 0.1 & \textcolor{red}{0.9125} \\
& Environmental Protection & -0.33 & \textcolor{red}{0.7418} \\
\midrule

Projects Proportionality & Education & -9.93 & 1.67 $\times$ 10 \textsuperscript{-20} \\
& Public Transit and Roads & 6.18 & 1.95 $\times$ 10 \textsuperscript{-9} \\
& Health & -1.19 & \textcolor{red}{0.23} \\
& Welfare & -5.57 & 8.83 $\times$ 10 \textsuperscript{-8} \\
& Public Space & 7.62 & 2.68 $\times$ 10 \textsuperscript{-13} \\
& Urban Greenery & 10.58 & 2.22 $\times$ 10 \textsuperscript{-22} \\
& Culture & -10.88 & 1.17 $\times$ 10 \textsuperscript{-23} \\
& Sport & 1.60 & 0.11 $\times$ 10 \textsuperscript{-23} \\
& Environmental Protection & 8.64 & 4.42 $\times$ 10 \textsuperscript{-16} \\
\midrule

Popularity Proportionality & Education & -10.85 & 1.19 $\times$ 10 \textsuperscript{-23} \\
& Public Transit and Roads & 4.42 & 1.35 $\times$ 10 \textsuperscript{-5} \\
& Health & -3.69 & 2.60 $\times$ 10 \textsuperscript{-4} \\
& Welfare & -7.74 & 5.95 $\times$ 10 \textsuperscript{-13} \\
& Public Space & 5.85 & 1.18 $\times$ 10 \textsuperscript{-8} \\
& Urban Greenery & 7.75 & 1.44 $\times$ 10 \textsuperscript{-13} \\
& Culture & -12.12 & 4.23 $\times$ 10 \textsuperscript{-28} \\
& Sport & 0.53 & \textcolor{red}{0.59} \\
& Environmental Protection & 5.60 & 5.05 $\times$ 10 \textsuperscript{-8} \\
\bottomrule
\end{tabular}
\end{scriptsize}
\end{table}

% Tabular data for t-test results on impact metrics at ballot level
\begin{table}[!htb]
\centering
% \captionsetup{justification=raggedright, singlelinecheck=false}
\caption{Paired t-test statistics for impact assessment at the voters' ballots level for share, representation and proportionality metrics}
\label{paired_t_test_impact_metrics_ballot_level}

\begin{scriptsize}

\begin{tabular}{cccc}
\toprule
Impact Metrics & Impact Areas & t-statistic & p-value \\
\midrule
Cost Share & Education & -1.76 & \textcolor{red}{0.0792} \\
& Public Transit and Roads & 6.14 & 2.56 $\times$ 10 \textsuperscript{-9} \\
& Health & 3.76 & 0.0002 \\
& Welfare & -2.32 & 0.0218 \\
& Public Space & 8.68 & 1.83 $\times$ 10 \textsuperscript{-16} \\
& Urban Greenery & 6.42 & 5.34 $\times$ 10 \textsuperscript{-10} \\
& Culture & -4.29 & 2.35 $\times$ 10 \textsuperscript{-5} \\
& Sport & 6.38 & 5.98 $\times$ 10 \textsuperscript{-10} \\
& Environmental Protection & 6.53 & 3.17 $\times$ 10 \textsuperscript{-10} \\
\midrule
Projects Share & Education & 0.74 & \textcolor{red}{0.4580} \\
& Public Transit and Roads & 7.35 & 1.86 $\times$ 10 \textsuperscript{-12} \\
& Health & 5.69 & 3.02 $\times$ 10 \textsuperscript{-8} \\
& Welfare & -1.22 & \textcolor{red}{0.22} \\
& Public Space & 12.03 & 6.69 $\times$ 10 \textsuperscript{-28} \\
& Urban Greenery & 10.19 & 4.73 $\times$ 10 \textsuperscript{-21} \\
& Culture & -1.39 & \textcolor{red}{0.1629} \\
& Sport & 5.81 & 1.53 $\times$ 10 \textsuperscript{-8} \\
& Environmental Protection & 10.74 & 1.25 $\times$ 10 \textsuperscript{-22} \\
\midrule
Cost Representation & Education & -3.05 & 0.0024 \\
& Public Transit and Roads & 5.37 & 1.55 $\times$ 10 \textsuperscript{-7}\\
& Health & 0.37 & \textcolor{red}{0.7140} \\
& Welfare & -5.87 & 2.35 $\times$ 10 \textsuperscript{-8}  \\
& Public Space & 6.81 & 4.47 $\times$ 10 \textsuperscript{-11} \\
& Urban Greenery & 5.42 & 1.23 $\times$ 10 \textsuperscript{-7} \\
& Culture & -6.58 & 1.99 $\times$ 10 \textsuperscript{-10} \\
& Sport & 4.39 & 1.49 $\times$ 10 \textsuperscript{-5} \\
& Environmental Protection & 3.83 & 0.0002 \\
\midrule
Projects Representation & Education & -11.53 & 4.68 $\times$ 10 \textsuperscript{-26} \\
& Public Transit and Roads & 0.33 & \textcolor{red}{0.7414} \\
& Health & -5.44 & 1.10 $\times$ 10 \textsuperscript{-7} \\
& Welfare & -8.29 & 3.55 $\times$ 10 \textsuperscript{-14} \\
& Public Space & -0.74 & \textcolor{red}{0.4584} \\
& Urban Greenery & 1.82 & \textcolor{red}{0.0699} \\
& Culture & -13.63 & 1.63 $\times$ 10 \textsuperscript{-33} \\
& Sport & -2.26 & 0.0239 \\
& Environmental Protection & -0.13 & \textcolor{red}{0.8937} \\
\midrule
Cost Proportionality & Education & -4.10 & 5.13 $\times$ 10 \textsuperscript{-5} \\
& Public Transit and Roads & 4.57 & 7.02 $\times$ 10 \textsuperscript{-6} \\
& Health & -1.48 & \textcolor{red}{0.1406} \\
& Welfare & -5.63 & 7.69 $\times$ 10 \textsuperscript{-8} \\
& Public Space & 5.10 & 5.71 $\times$ 10 \textsuperscript{-7} \\
& Urban Greenery & 3.78 & 0.0002 \\
& Culture & -5.47 & 9.37 $\times$ 10 \textsuperscript{-8} \\
& Sport & -3.64 & 0.0003 \\
& Environmental Protection & 2.87 & 0.0045 \\
\midrule
Projects Proportionality & Education & 0.19 & \textcolor{red}{0.8483} \\
& Public Transit and Roads & 9.39 & 1.47 $\times$ 10 \textsuperscript{-18} \\
& Health & 4.54 & 8.02 $\times$ 10 \textsuperscript{-6} \\
& Welfare & -1.16 & \textcolor{red}{0.2473}  \\
& Public Space & 10.16 & 2.73 $\times$ 10 \textsuperscript{-21} \\
& Urban Greenery & 12.38 & 1.64 $\times$ 10 \textsuperscript{-28}  \\
& Culture & -1.76 & \textcolor{red}{0.0790} \\
& Sport & 6.57 & 1.94 $\times$ 10 \textsuperscript{-10}\\
& Environmental Protection & 10.16 & 9.66 $\times$ 10 \textsuperscript{-21} \\
\bottomrule
\end{tabular}
\end{scriptsize}
\end{table}

% Tabular data for t-test results on novelty metrics
\begin{table}[!htb]
\centering
\captionsetup{justification=raggedright, singlelinecheck=false}
\caption{Paired t-test statistics for novelty assessment at the voting outcome level across different impact areas}
\label{paired_t_test_novelty_metrics}
\begin{scriptsize}
\begin{tabular}{cccc}

\toprule
Novelty Metrics & Impact Areas & t-statistic & p-value \\
\midrule
Cost Novelty {\em Within} Impact Areas & Education & -13.43 & 1.14 $\times$ 10 \textsuperscript{-32} \\
 & Public Transit and Roads & 0.40 & \textcolor{red}{0.69} \\
& Health & -4.35 & 1.99 $\times$ 10 \textsuperscript{-5} \\
& Welfare & -6.49 & 1.99 $\times$ 10 \textsuperscript{-9} \\
& Public Space & 2.51 & 0.01 \\
& Urban Greenery & 1.74 & \textcolor{red}{0.08} \\
& Culture & -15.81 & 6.98 $\times$ 10 \textsuperscript{-41} \\
& Sport & -1.69 & \textcolor{red}{0.09} \\
& Environmental Protection & 1.11 & \textcolor{red}{0.27} \\
\midrule
Projects Novelty {\em Within} Impact Areas & Education & -20.53 & 1.24 $\times$ 10 \textsuperscript{-59} \\
 & Public Transit and Roads & -6.20 & 2.02 $\times$ 10 \textsuperscript{-9} \\
& Health & -10.61 & 8.58 $\times$ 10 \textsuperscript{-22} \\
& Welfare & -8.29 & 5.19 $\times$ 10 \textsuperscript{-13} \\
& Public Space & -12.79 & 1.19 $\times$ 10 \textsuperscript{-30} \\
& Urban Greenery & -6.49 & 3.96 $\times$ 10 \textsuperscript{-10} \\
& Culture & -19.99 & 3.54 $\times$ 10 \textsuperscript{-56} \\
& Sport & -10.91 & 1.49 $\times$ 10 \textsuperscript{-23} \\
& Environmental Protection & -7.29 & 4.12 $\times$ 10 \textsuperscript{-12} \\
\midrule
Popularity Novelty {\em Within} Impact Areas & Education & -17.55 & 2.76 $\times$ 10 \textsuperscript{-48} \\
 & Public Transit and Roads & -2.56 & 0.0109 \\
& Health & -7.71 & 3.37 $\times$ 10 \textsuperscript{-13} \\
& Welfare & -7.15 & 1.42 $\times$ 10 \textsuperscript{-10} \\
& Public Space & -6.82 & 4.53 $\times$ 10 \textsuperscript{-11} \\
& Urban Greenery & -2.63 & 0.0091 \\
& Culture & -17.53 & 3.48 $\times$ 10 \textsuperscript{-47} \\
& Sport & -7.01 & 1.58 $\times$ 10 \textsuperscript{-11} \\
& Environmental Protection & -3.38 & 0.0008 \\
\midrule
Cost Novelty {\em Between} Impact Areas & Education & -10.31 & 2.58 $\times$ 10 \textsuperscript{-21} \\
 & Public Transit and Roads & 5.38 & 1.59 $\times$ 10 \textsuperscript{-7} \\
& Health & -0.77 & \textcolor{red}{0.44} \\
& Welfare & -5.36 & 1.76 $\times$ 10 \textsuperscript{-7} \\
& Public Space & 6.88 & 3.96 $\times$ 10 \textsuperscript{-11} \\
& Urban Greenery & 6.61 & 1.94 $\times$ 10 \textsuperscript{-10} \\
& Culture & -11.68 & 6.59 $\times$ 10 \textsuperscript{-26} \\
& Sport & 4.35 & 1.93 $\times$ 10 \textsuperscript{-5} \\
& Environmental Protection & 5.31 & 2.27 $\times$ 10 \textsuperscript{-7} \\
\midrule
Projects Novelty {\em Between} Impact Areas & Education & -12.94 & 2.86 $\times$ 10 \textsuperscript{-30} \\
 & Public Transit and Roads & 6.61 & 2.01 $\times$ 10 \textsuperscript{-10} \\
& Health & 0.44 & \textcolor{red}{0.66} \\
& Welfare & -5.69 & 3.18 $\times$ 10 \textsuperscript{-8} \\
& Public Space & 9.14 & 1.37 $\times$ 10 \textsuperscript{-17} \\
& Urban Greenery & 8.11 & 1.62 $\times$ 10 \textsuperscript{-14} \\
& Culture & -13.66 & 7.94 $\times$ 10 \textsuperscript{-33} \\
& Sport & 4.57 & 7.33 $\times$ 10 \textsuperscript{-6} \\
& Environmental Protection & 5.43 & 1.25 $\times$ 10 \textsuperscript{-7} \\
\midrule
Popularity Novelty {\em Between} Impact Areas & Education & -11.85 & 1.71 $\times$ 10 \textsuperscript{-26} \\
 & Public Transit and Roads & 6.09 & 3.75 $\times$ 10 \textsuperscript{-9} \\
& Health & 0.45 & \textcolor{red}{0.65} \\
& Welfare & -5.42 & 1.29 $\times$ 10 \textsuperscript{-7} \\
& Public Space & 8.08 & 2.06 $\times$ 10 \textsuperscript{-14} \\
& Urban Greenery & 7.42 & 1.40 $\times$ 10 \textsuperscript{-12} \\
& Culture & -12.93 & 3.06 $\times$ 10 \textsuperscript{-30} \\
& Sport & 4.64 & 5.27 $\times$ 10 \textsuperscript{-6} \\
& Environmental Protection & 5.21 & 3.68 $\times$ 10 \textsuperscript{-7} \\
\bottomrule
\end{tabular}
\end{scriptsize}
\end{table}

% table for conjoint analysis p-values
\begin{table}[!htb]
\centering
\caption{Relative importance and p-values for different project tag combination used in conjoint analysis of budget utilization rate}
\label{conjoint_analysis_results}
\scriptsize
\begin{tabular}{ccccccc}
\toprule
\multicolumn{2}{c}{Tag Combination} & Cost Level& \multicolumn{2}{c}{p-value}& \multicolumn{2}{c}{Relative Importance} \\
\cmidrule(lr){4-5} \cmidrule(lr){6-7}
& & & UG& ES& UG& ES\\
\midrule
\multicolumn{2}{c}{Culture, Education} & Low & $5.84 \times 10 \textsuperscript{-32}$ & $8.94 \times 10 \textsuperscript{-56}$ & $0.52$ & $0.71$ \\
\multicolumn{2}{c}{Culture, Education} & High & $0.002$ & $0.03$ & $-0.17$ & $-0.16$ \\
\multicolumn{2}{c}{Env. Protection, Public Space, Urban Greenery} &  Low& $6.48 \times 10 \textsuperscript{-8}$& $0.0004$& $0.05$& $-0.06$ \\
\multicolumn{2}{c}{Env. Protection, Public Space, Urban Greenery} &  High& $2.33 \times 10 \textsuperscript{-9}$& $0.03$& $0.11$& $-0.15$ \\
\multicolumn{2}{c}{Health, Public Space, Sport} &  Low& $3.38 \times 10 \textsuperscript{-7}$& $1.09 \times 10 \textsuperscript{-8}$& $0.14$& $0.10$ \\
\multicolumn{2}{c}{Health, Public Space, Sport} &  High& $\textcolor{red}{0.19}$& $0.001$& $-0.47$& $-0.28$ \\
\multicolumn{2}{c}{Public Space, Public Transit and Roads} &  Low& $1.79 \times 10 \textsuperscript{-5}$& $1.44 \times 10 \textsuperscript{-5}$& $-0.04$& $-0.02$ \\
\multicolumn{2}{c}{Public Space, Public Transit and Roads} &  High& $3.37 \times 10 \textsuperscript{-4}$& $0.01$& $-0.13$& $-0.13$ \\
\bottomrule
\end{tabular} 
\end{table}

\clearpage
\makeatletter\@input{xx.tex}\makeatother